\documentclass[useAMS,usenatbib]{mn2e}
\usepackage[T1]{fontenc}
\usepackage{aecompl}

\usepackage{amsmath,enumitem,booktabs,fixltx2e,mdframed}
\usepackage{psfig,graphicx,threeparttable,xspace}
\newcommand{\f}{\ensuremath{f_{\mathrm{s}}}\xspace}
\newcommand{\vsini}{\ensuremath{v\sin{i}}\xspace}
\newcommand{\kms}{kms\ensuremath{^{-1}}\xspace}
\newcommand{\kb}{\ensuremath{K_{2}}\xspace}
\newcommand{\prot}{\ensuremath{P_{\text{rot}}}\xspace}

\usepackage{xfrac,dcolumn}
\usepackage{float}

\newcount\longrefs
\def\aap{\ifnum\longrefs=1 {Astron.\ Astrophys.}\else 
                           {A\hbox{\rm \&}A}\fi}
\def\aapr{\ifnum\longrefs=1 {Astron.\ Astrophys.\ Rev.}\else 
                            {A\hbox{\rm \&}AR}\fi}
\def\aaps{\ifnum\longrefs=1 {Astron.\ Astrophys.\ Suppl.}\else 
                            {A\hbox{\rm \&}A Suppl.}\fi}
\def\aipcs{\ifnum\longrefs=1 {Am.\ Inst.\ Phys.\ Conf.\ Series}\else
                             {AIP Conf.\ Ser.}\fi}
\def\aj{\ifnum\longrefs=1 {Astron.\ J.}\else 
                          {AJ}\fi} 
\def\ao{\ifnum\longrefs=1 {Applied Optics}\else 
                           {Appl.\ Opt.}\fi} 
\def\aspcs{\ifnum\longrefs=1 {Astron.\ Soc.\ Pacific Conf.\ Series}\else 
                           {ASP Conf.\ Ser.}\fi} 
\def\apj{\ifnum\longrefs=1 {Astrophys.\ J.}\else 
                           {ApJ}\fi} 
\def\apjl{\ifnum\longrefs=1 {Astrophys.\ J. Lett.}\else 
                            {ApJ}\fi} 
\def\aplett{\ifnum\longrefs=1 {Astrophys.\ J. Lett.}\else 
                            {ApJ}\fi} 
\def\apjs{\ifnum\longrefs=1 {Astrophys.\ J. Suppl.}\else 
                            {ApJS}\fi}
\def\apss{\ifnum\longrefs=1 {Astrophys.\ and Space Science}\else 
                            {Astrophys.\ Space Sci.}\fi}
\def\araa{\ifnum\longrefs=1 {Ann.\ Rev.\ Astron.\ Astrophys.}\else 
                            {ARA\hbox{\rm \&}A}\fi}
\def\azh{\ifnum\longrefs=1 {Astronomicheskii Zhurnal}\else 
                            {Astron.\ Zhur.}\fi}
\def\baas{\ifnum\longrefs=1 {Bull.\ Am.\ Astron.\ Soc.}\else 
                            {BAAS}\fi}
\def\bain{\ifnum\longrefs=1 {Bull.\ Astronom.\ Institutes Netherlands}\else
                            {Bull.\ Astr.\ Inst.\ Neth.}\fi}
\def\gca{\ifnum\longrefs=1 {Geochim.\ Cosmochim.\ Acta}\else 
                           {Geochim.\ Cosmochim.\ Acta}\fi}
\def\grl{\ifnum\longrefs=1 {Geophys.\ Res.\ Lett.}\else 
                           {Geoph.\ Res.\ Lett.}\fi}
\def\iaucirc{\ifnum\longrefs=1 {IAU Circulars}\else 
                          {IAU Circ.}\fi}
\def\ip{\ifnum\longrefs=1 {in press}\else 
                          {in press}\fi}
\def\jgr{\ifnum\longrefs=1 {J.\ Geophys.\ Res.}\else 
                           {J.\ Geophys.\ Res.}\fi}  
\def\jrasc{\ifnum\longrefs=1 {J.\ Royal Astron.\ Soc.\ Canada}\else 
                           {JRAS Can.}\fi}  
\def\memsai{\ifnum\longrefs=1 {Mem.~Soc.~Astron.~Italiana}\else
                              {MemSAI}\fi}
\def\mnras{\ifnum\longrefs=1 {Mon.\ Not.\ Roy.\ Astron.\ Soc.}\else 
                             {MNRAS}\fi} 
\def\nat{\ifnum\longrefs=1 {Nature}\else 
                           {Nat}\fi}
\def\pasj{\ifnum\longrefs=1 {Pub.\ Astron.\ Soc.\ Japan}\else 
                            {PASJ}\fi} 
\def\pasp{\ifnum\longrefs=1 {Pub.\ Astron.\ Soc.\ Pacific}\else 
                            {PASP}\fi} 
\def\physscr{\ifnum\longrefs=1 {Physica Scripta}\else 
                            {Phys.\ Scrip.}\fi} 
\def\planss{\ifnum\longrefs=1 {Planetary \& Space Science}\else 
                            {Plan. \& Space Sci.}\fi} 
\def\procspie{\ifnum\longrefs=1 {Proc.\ SPIE}\else 
                            {Proc.\ SPIE}\fi} 
\def\qjras{\ifnum\longrefs=1 {Quarterly J.\ Royal Astron.\ Soc.}\else 
                            {QJRAS}\fi} 
\def\sa{\ifnum\longrefs=1 {Soviet Astron..}\else 
                               {Sov.\ Astron.}\fi}
\def\skytel{\ifnum\longrefs=1 {Sky \& Telescope}\else 
                            {Sky \& Tel.}\fi} 
\def\solphys{\ifnum\longrefs=1 {Solar Phys.}\else 
                               {Sol.\ Phys.}\fi}
\def\ssr{\ifnum\longrefs=1 {Space Science Rev.}\else 
                               {Space\ Sci.\ Rev.}\fi}
\def\zap{\ifnum\longrefs=1 {Zeitschr.\ f.\ Astrophysik}\else
                               {Z.\ Astrophys.}\fi}

\title[The long-term magnetic activity of AE Aqr]{Roche tomography of cataclysmic variables - VII. \\The long-term magnetic activity of AE Aqr}
\author[C.A. Hill et  al.]{C.A. Hill$^{1,2}$\thanks{E-mail:
chill17@qub.ac.uk}, C.A. Watson$^{1}$, D. Steeghs$^{3}$, V.S. Dhillon$^{4,5}$ and T. Shahbaz$^{5,6}$\\
$^1$Astrophysics Research Centre, Queen's University Belfast, Belfast, BT7 1NN, Northern Ireland, UK\\
$^2$IRAP, Observatoire Midi-Pyr\'{e}n\'{e}es, University of Toulouse, 14 avenue Edouard Belin, 31400, Toulouse, France\\
$^3$Department of Physics, University of Warwick, Coventry, CV4 7AL, UK\\
$^4$Department of Physics \& Astronomy, University of Sheffield, Sheffield, S3 7RH, UK\\
$^5$Instituto de Astrof\'{i}sica de Canarias (IAC), E-38200 La Laguna, Tenerife, Spain \\
$^6$Departamento de Astrof\'{i}sica, Universidad de La Laguna (ULL), E-38206 La Laguna, Tenerife, Spain}
\begin{document}

\date{}

\pagerange{\pageref{firstpage}--\pageref{lastpage}} \pubyear{2015}

\maketitle

\label{firstpage}

\begin{abstract}
We present a long-term study of the secondary star in the cataclysmic variable AE~Aqr, using Roche tomography to indirectly image starspots on the stellar surface spanning 8~years of observations. The 7 maps show an abundance of spot features at both high and low latitudes. We find that all maps have at least one large high-latitude spot region, and we discuss its complex evolution between maps, as well as its compatibility with current dynamo theories. Furthermore, we see the apparent growth in fractional spot coverage, \f, around $45\degr$~latitude over the duration of observations, with a persistently high \f near latitudes of $20\degr$. These bands of spots may form as part of a magnetic activity cycle, with magnetic flux tubes emerging at different latitudes, similar to the `butterfly' diagram for the Sun. We discuss the nature of flux tube emergence in close binaries, as well as the activity of AE~Aqr in the context of other stars.
\end{abstract}


\section{Introduction}
\label{sec:intro}
Understanding the behaviour of stellar magnetic activity, and the nature of the underlying dynamo mechanism, are some of the most pressing challenges in solar and stellar physics. It is well known that the Sun displays an 11~yr sun spot cycle. Since the first detection of cyclic magnetic behaviour in solar-like stars (e.g. \citealt{wilson1978}), there has been great interest in determining which parameters, such as binarity, spin, or convective zone depth (and hence stellar type), are pivotal to both the duration and amplitude of magnetic activity cycles. 

In a survey of stellar activity on 111 lower main-sequence stars, \cite{baliunas1995} used chromospheric Ca~\textsc{ii}~HK measurements as a proxy of the surface magnetic fields. They found that, of the stars with solar-like activity cycles, the measured activity cycle periods $P_{\mathrm{cyc}}$ ranged from 2.5~yr to the 25~yr maximum baseline of observations. They also found that G0--K5V type stars show changes in rotation and chromospheric activity on an evolutionary timescale, with stars of intermediate age showing moderate levels of activity and occasional smooth cycles, whereas young rapidly-rotating stars exhibit high average levels of activity and rarely display a smooth, cyclic variation.

In other work, \cite{saar1999} used a large and varied stellar sample (including evolved stars and cataclysmic variable secondaries) to explore the relationships between the length of the activity cycle $P_{\text{cyc}}$ and the stellar rotation period $P_{\text{rot}}$. They parameterized the relationships using the ratio of cycle and rotation frequencies ${\omega_{\text{cyc}}/\Omega~(=P_{\text{rot}}/P_{\text{cyc}})}$, as well as the inverse Rossby number ${\text{Ro}^{-1}~(\equiv 2 \tau_{\text{c}} \Omega}$, where $\tau_{\text{c}}$ is the convective turnover timescale). They found that stars with ages >0.1~Gyr lay on two nearly parallel branches, separated by a factor of $\sim6$ in ${\omega_{\text{c}}/\Omega}$, with both branches exhibiting increasing ${\omega_{\text{c}}/\Omega}$ with increasing $\text{Ro}^{-1}$. Furthermore, they found that, if the secondary stars in close binaries can be used as proxies for young, rapidly rotating single stars, the cycles of these stars populate a third `superactive' branch, that shows the opposite trend of \emph{decreasing} ${\omega_{\text{cyc}}/\Omega}$ with increasing $\text{Ro}^{-1}$.

Elsewhere, \cite{radick1998} found that the luminosity variation of young stars was anti-correlated with their chromospheric emission, in the sense that young stars are fainter near their activity maxima. This suggests that the long-term variability of young stars is spot-dominated, whereas older stars are faculae-dominated (\citealt{lockwood2007}). Such behaviour has been observed in a number of young single stars (e.g. \citealt{berdyugina2002,messina2002,jarvinen2005}) as well as in binary systems (e.g. \citealt{henry1995}). Indeed, magnetic activity cycles have been found in several systems using photometric techniques, with some systems appearing to show preferred longitudes for spot activity. The increase and corresponding decrease of spot activity on opposite stellar longitudes has been interpreted as a so-called `flip-flop' magnetic activity cycle (e.g. \citealt{berdyugina1998,berdyugina2005flipflop}). By tracking the number and position of spots on the stellar surface using Doppler imaging, such activity was also found on the RS~CVn star, II~Peg, by \cite{berdyugina1999}. 

While the magnetic activity of single stars and detached binaries is reasonably well studied, studies of magnetic activity cycles on interacting binaries are critically lacking. Cataclysmic variables (CVs) are semi-detached binaries consisting of a (typically) lower main-sequence star transferring mass to a white dwarf (WD) primary via Roche-lobe overflow. These systems, with both rapid rotation and tidal distortion, provide a unique parameter regime to allow critical tests of stellar dynamo theories. In addition, CVs form the foundation of our understanding of a wide range of accretion driven phenomena, and in turn, the secondary stars are key to our understanding of the origin, evolution and behaviour of this class of interacting binary. The secondary star regulates the mass transfer history and is intimately tied in with the orbital angular momentum transport that determines the evolutionary timescales of the various accretion stages. In particular, magnetic braking is thought to drain angular momentum from the system, sustaining the mass transfer that causes CVs to evolve to shorter orbital periods. This has been a standard ingredient of compact binary evolution theory for several decades. 

Furthermore, magnetic activity cycles in secondary stars have been invoked to explain the variations in orbital periods in interacting binaries caused by the \cite{applegate1992} mechanism. This causes angular momentum changes within the secondary star throughout the activity cycle to be transmitted to the orbital motion, resulting in cyclical orbital period variations. In addition, an increase in the number of magnetic flux tubes on the secondary star during a stellar maximum is thought to cause the star to expand (\citealt{richman1994}) and to result in enhanced mass transfer -- giving rise to an increased mass transfer rate through the disc and a corresponding increase in the system luminosity. Additional mass transfer also reduces the time required to build up sufficient material in the disc to trigger an outburst, resulting in shorter time intervals between consecutive outbursts (e.g. \citealt{bianchini1990}). On shorter timescales, starspots are thought to quench mass transfer from the secondary star as they pass the mass-losing `nozzle', resulting in the low-states observed in many CVs (see \citealt{livio1994,king1998,hessman2000}). Previous surface maps of the secondary stars in the CVs BV~Cen (\citealt{watson2007}) and AE~Aqr (\citealt{watson2006,hill2014}) show a dramatic increase in spot coverage on the side of the star facing the WD. This suggests that magnetic flux tubes are forced to emerge at preferred longitudes, as predicted by \cite{holzwarth2003b}, and is possibly related to the impact of tidal forces from the nearby compact object. If these particular spot distributions are confirmed to be long-lasting features, they would require explanation by stellar dynamo theory (e.g. \citealt{sokoloff2002,moss2002}), and would provide evidence for the impact of tidal forces on magnetic flux emergence. In addition, since the number of star spots should change dramatically over the course of an activity cycle, the density of spots around the mass transfer nozzle may also vary. This would provide an explanation for the extended high and low periods seen in polar type CVs such as AM Her (\citealt{hessman2000}).

Thus, the magnetic activity of CV secondary stars is crucial to the long and short term behaviour of these systems. Furthermore, it is clear that comparisons of the long-term magnetic activity across a range of stellar types, in different systems, are crucial to understanding the nature of the stellar dynamo, how it evolves, and what system parameters are most important in its operation. In light of this, we present a study of the long-term magnetic activity of the secondary star in the CV, AE~Aqr ($\prot = 9.88$~h) by using Roche tomography to map the number, size, distribution and variability of starspots on the surface. This is the first time a CV secondary has been tracked with this type of a campaign, and given that CVs with both rapid-rotation and tidal distortion provide unique test-beds for dynamo theories, we can better understand what parameters are most important to the behaviour of the underlying dynamo mechanism, and the duration and amplitude of magnetic activity cycles

\section{Observations and reduction}
Simultaneous spectroscopic and photometric data of AE Aqr were taken in 2001, 2004, 2005 and 2006 (hereafter D01, D04, D05, D06), with spectroscopic data only in 2008 and 2009 (hereafter D08, D09a, D09b), where D9a and D09b were taken 9~d apart. As D01 and D09a~\&~D09b have previously been published in \cite{watson2006} and \cite{hill2014}, respectively, we refer the reader to these works for details of the reduction methods for both the spectroscopic and photometric data. Logs of the observations are shown in Tables~\ref{tab:obs}~and~\ref{tab:obsphotometry}. 

\subsection{Spectroscopy}
\subsubsection{MIKE+Magellan Clay telescope}
For D04, D05 and D06, spectroscopic observations were carried out using the dual-beam Magellan Inamori Kyocera Echelle spectrograph (MIKE, \citealt{bernstein2003}) on the 6.5 m Magellan Clay telescope, situated at the Las Campanas Observatory in Chile. The standard set-up was used, allowing a wavelength coverage of 3330--5070 \r{A} in the blue arm and 4460--7270 \r{A} in the red arm, with significant wavelength overlap between adjacent orders. A slit width of 0.7 arcsec was used, providing a spectral resolution of around 38,100 ($\sim7.8$~\kms) and 31,500 ($\sim9.5$~\kms) in the blue and red channels, respectively. A Gaussian fit to several arc lamp lines gave a mean instrumental resolution of $\sim9$~\kms, which was adopted for use in Roche tomography in Section~\ref{sec:rochetomography}.  Exposure times of 250~s (0.7~per~cent of the orbital period) were used in order to minimize velocity smearing of the data due to the orbital motion of the secondary star. ThAr lamp exposures were taken every 10 exposures for the purpose of wavelength calibration.

The data were reduced using the MIKE pipeline written in \textsc{python} by \cite{mikepipeline}. This automatically conducts bias subtraction, flat-fielding, blaze correction and wavelength calibration. The final output provides 1-D spectra split into orders, for both the blue and red arms. After reduction, it was found that each extracted order was not fully blaze-corrected, and so we applied an additional correction using a flux standard star. After flux calibration, the orders in the blue and red arms, respectively, were combined into continuous spectra by taking a variance-weighted mean across the spectral range. The blue spectra were then scaled to match the red spectra by optimally-subtracting the overlapping spectral regions (where the blue spectra was scaled and subtracted from the red spectra, with the optimal scaling factor being that which minimizes the residuals). Finally, a variance-weighted mean was made by combining the spectra from both arms, creating a single spectrum for each exposure.

\subsubsection{UVES+VLT}
For D08, spectroscopic observations were carried out using the Ultraviolet and Visual Echelle Spectrograph (UVES, \citealt{dekker2000}) on the 8.2-m UT2 of the VLT, situated on Cerro Paranal in Chile. UVES was used in the Dichroic-1/Standard setting (390+580 nm) mode, allowing a wavelength coverage of $3282-4563$~\r{A} in the blue arm and $4726-6835$~\r{A} in the red arm. A slit width of 0.9~arcsec was used, providing a spectral resolution of around 46,000 ($\sim6.5$~\kms) and 43,000 ($\sim7$~\kms) in the blue and red channels, respectively -- an instrumental resolution of 7~\kms was adopted for use in Roche tomography in Section~\ref{sec:rochetomography}. Exposure times of 230~s (0.65~per~cent of the orbital period) were used, with ThAr lamp exposures taken at the start and end of the night. The data were taken from the European Southern Observatory (ESO) data products archive after being reduced automatically using version 5.1.5 of the ESO/UVES pipeline. The final output consisted of 1-D spectra for both the red and blue arms.

\subsection{Photometry}
\label{sec:photometry}
\begin{figure}
\centering
\includegraphics[width=0.5\textwidth]{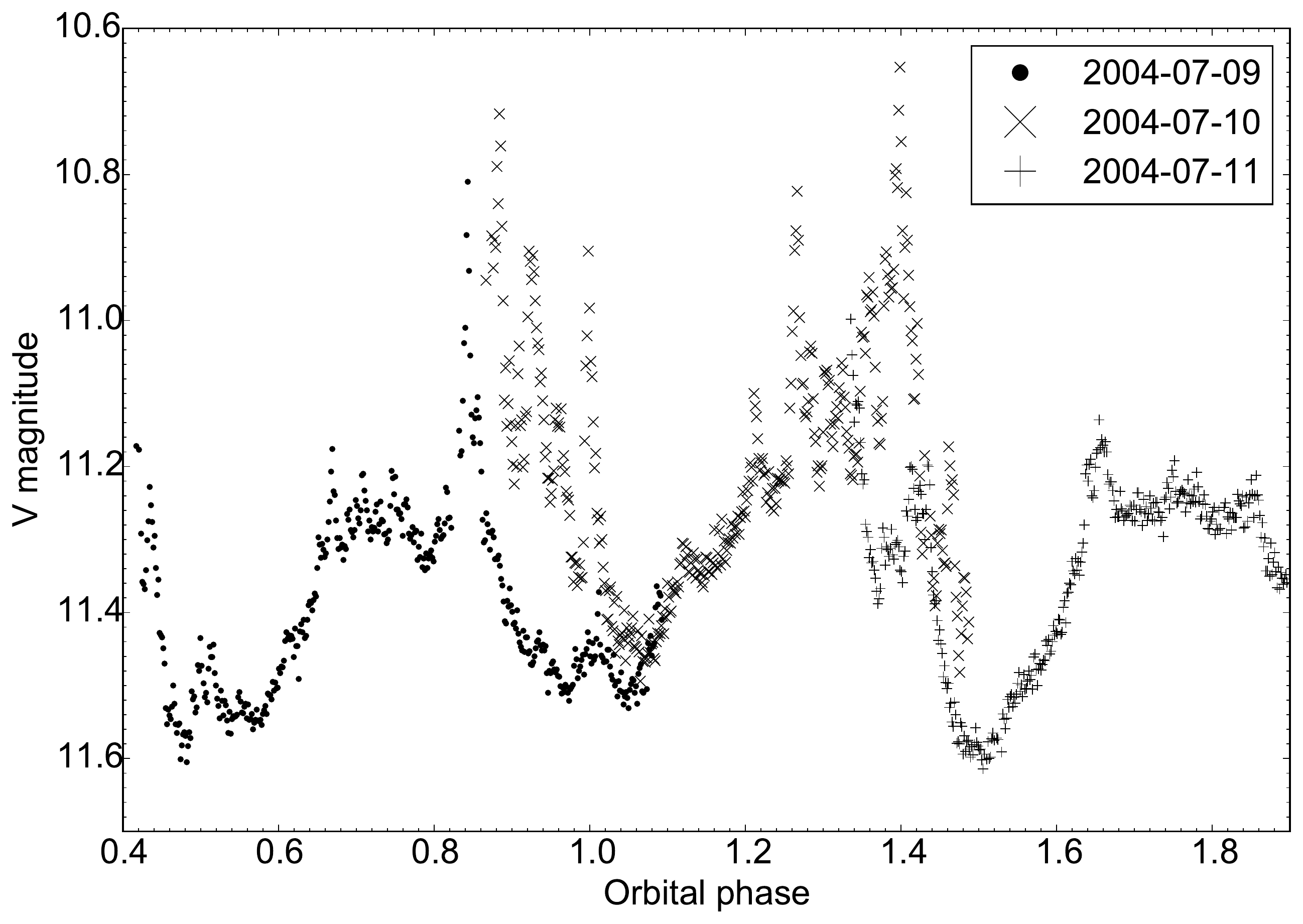}
\includegraphics[width=0.5\textwidth]{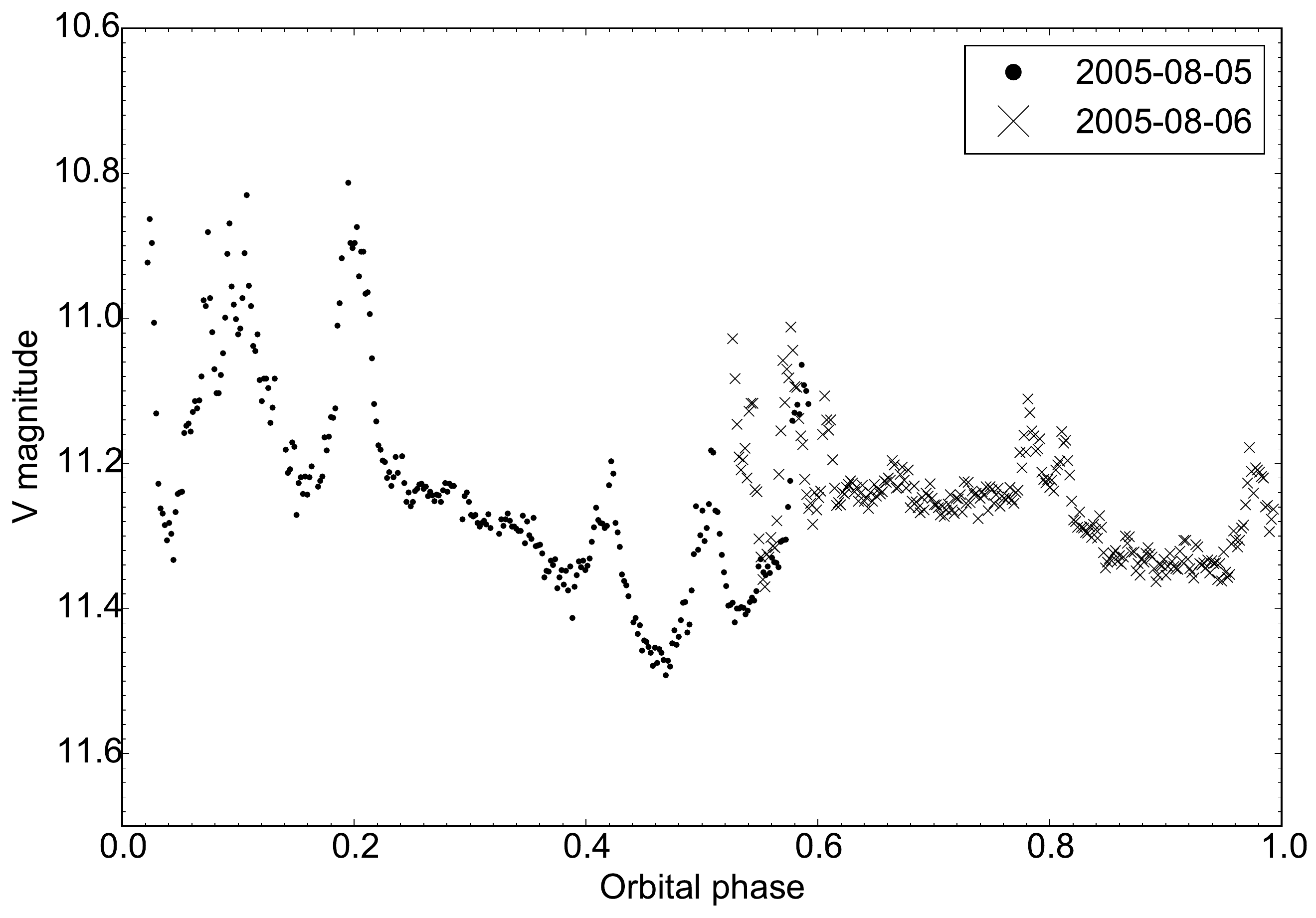}
\includegraphics[width=0.5\textwidth]{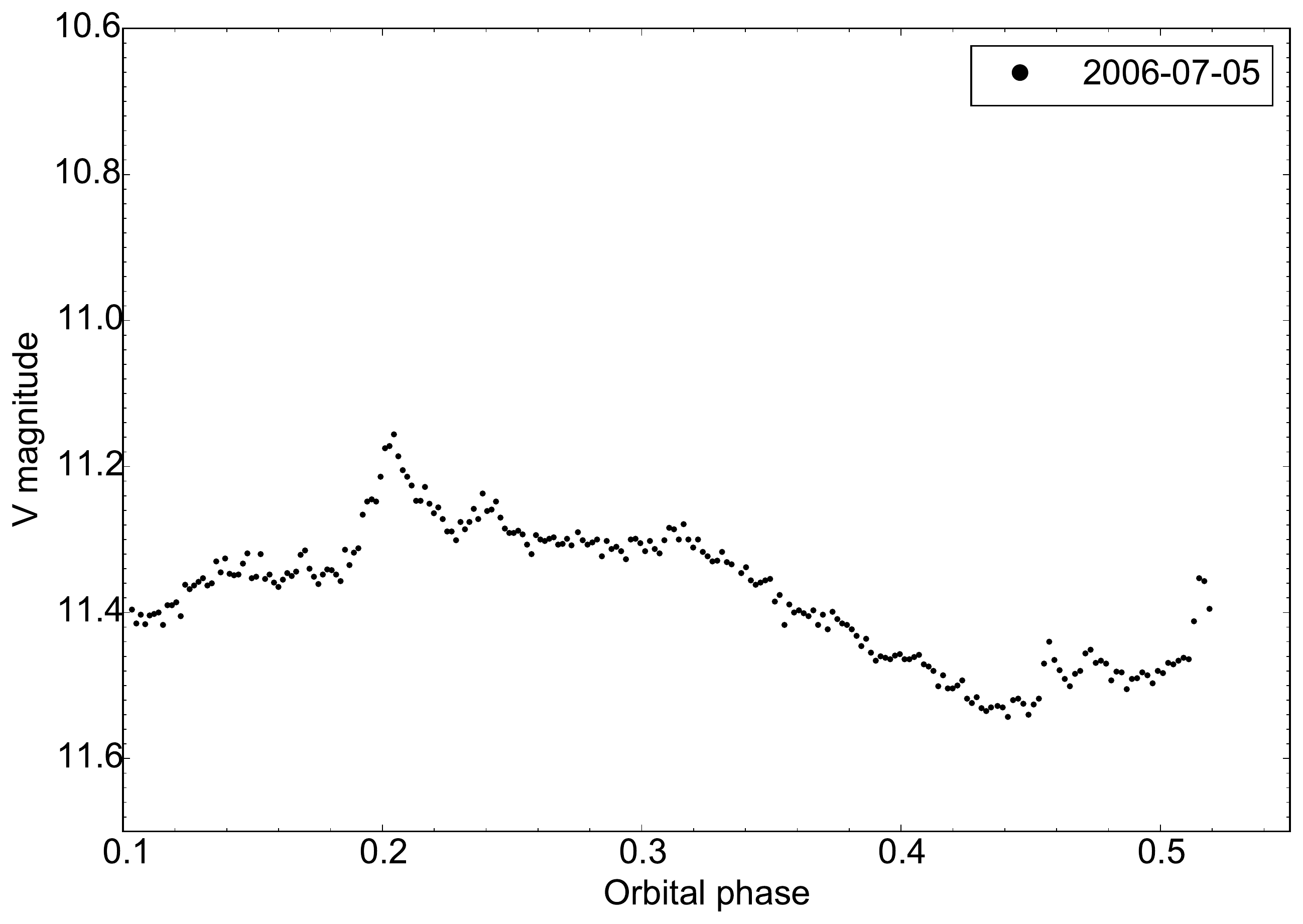}
\caption{The light curves of AE Aqr for D04 (top panel), D05 (middle) and D06 (bottom). The points are phase folded for clarity, and the typical uncertainties (not shown) are given in Table~\ref{tab:obsphotometry}. Rapid and frequent flaring is apparent in all plots, and is due to accretion variability.}
\label{fig:photometry}
\end{figure}

Simultaneous photometry was carried out for D04, D05 and D06 using a Harris V-band filter on the Carnegie Institution's Henrietta Swope 1-m telescope, situated at the Las Campanas Observatory in Chile. The data were reduced using standard techniques. The master bias frame showed no ramp or large scale structure across the CCD, and so the bias level of each frame was removed by subtracting the median value of pixels in the overscan region. Pixel-to-pixel variations were corrected by dividing the target frames by a master flat-field taken at twilight. Optimal photometry was performed using the package \textsc{photom} \citep{eaton2009}, where three suitable comparison stars were identified using the catalogue of \cite{henden1995} to perform differential photometry. The light curves of D04, D05 and D06 are shown in Figure~\ref{fig:photometry}. Flaring activity is clearly evident over both slow and rapid timescales, with amplitudes of up to $\sim0.6$~mag. This most likely stems from accretion variability rather than the secondary star.

\begin{table*}
\caption{A log of the spectroscopic observations of AE Aqr. Columns 1-3 list the UT date, the start, and end times of observations, respectively. Column 4 lists the instrument and telescope used. Columns 5-8 show the exposure time, the number of spectra taken, the peak signal-to-noise ratio around the central wavelength of each spectrum (with the typical value in parentheses), and the phase coverage achieved. Column 9 gives the abbreviation used throughout the text to refer to that specific data set.} 
\label{tab:obs}
\begin{tabular}{lcccccccc}
\toprule
\textsc{ut} date	&	\textsc{ut} start	&	\textsc{ut} end	&	Instrument	&	$T_{\mathrm{exp}}$~(s)	&	No. spectra	&	SNR	&	Phase coverage &	Abbreviation	\\
\midrule																	
2001 Aug 09	&	21:01	&	04:22	&	UES+WHT	&	200	&	88	&	$22-44$	&	$0.18-0.92$	&	D01	\\
2001 Aug 10	&	20:49	&	04:37	&	UES+WHT	&	200	&	95	&	$22-44$	&	$0.59-0.37$	&		\\
2004 Jul 09	&	02:42	&	09:18	&	MIKE+Magellan	&	250	&	69	&	$69-129 ~(\sim96)$	&	$0.46-0.81$	&	D04	\\
2004 Jul 10	&	03:08	&	08:57	&	MIKE+Magellan	&	250	&	64	&	$44-116 ~(\sim83)$	&	$0.91-0.49$	&		\\
2004 Jul 11	&	03:13	&	08:49	&	MIKE+Magellan	&	250	&	63	&	$64-121 ~(\sim81)$	&	$0.35-0.90$	&		\\
2005 Aug 05	&	01:49	&	07:33	&	MIKE+Magellan	&	250	&	64	&	$22-99 ~(\sim62)$	&	$0.03-0.60$	&	D05	\\
2005 Aug 06	&	02:49	&	07:29	&	MIKE+Magellan	&	250	&	57	&	$77-131 ~(\sim98)$	&	$0.56-0.02$	&		\\
2006 Jul 04	&	03:36	&	07:50	&	MIKE+Magellan	&	250	&	48	&	$45-100 ~(\sim95)$	&	$0.13-0.55$	&	D06	\\
2008 Aug 06	&	00:01	&	05:02	&	UVES+VLT	&	230	&	65	&	$91-185 ~(\sim125)$	&	$0.26-0.77$	&	D08	\\
2008 Aug 07	&	00:08	&	05:13	&	UVES+VLT	&	230	&	66	&	$119-176 ~(\sim149)$	&	$0.70-0.22$	&		\\
2009 Aug 27	&	23:49	&	05:11	&	UVES+VLT	&	230	&	57	&	$81-150 ~(\sim121)$	&	$0.34-0.89$	&	D09a	\\
2009 Aug 28	&	00:08	&	04:55	&	UVES+VLT	&	230	&	61	&	$76-147 ~(\sim129)$	&	$0.90-0.28$	&		\\
2009 Sept 05	&	23:43	&	04:38	&	UVES+VLT	&	230	&	61	&	$83-125 ~(\sim109)$	&	$0.19-0.68$	&	D09b	\\
2009 Sept 06	&	00:34	&	05:20	&	UVES+VLT	&	230	&	60	&	$107-158 ~(\sim136)$	&	$0.71-0.18$	&		\\
\bottomrule
\end{tabular}
\end{table*}

\begin{table*}
\caption{A log of the photometric observations taken of AE Aqr taken with the Henrietta Swope 1-m telescope. Columns 1-3 list the date, the start and end times of observations, respectively. Columns 4-6 give the exposure time, the number of exposures taken, and the typical uncertainty in the measured magnitude. Column 7 gives the abbreviation used throughout the text to refer to that specific data set.} 
\label{tab:obsphotometry}
\begin{tabular}{lcccccc}
\toprule
\textsc{ut} date	&	\textsc{ut} start	&	\textsc{ut} end	&	$T_{\mathrm{exp}}$ (s)	&	No. exp.	&	$\sigma_{mag}$ &	Abbreviation	\\
\midrule													
2004 Jul 09	&	02:42	&	09:22	&	10	&	416	&	$\sim0.023$ 	&	D04	\\
2004 Jul 10	&	02:54	&	09:02	&	10	&	384	&		"		&		\\
2004 Jul 11	&	03:14	&	08:50	&	10	&	348	&		"		&		\\
2005 Aug 05	&	01:55	&	07:33	&	20	&	290	&	$\sim0.024$ 	&	D05	\\
2005 Aug 06	&	02:56	&	07:32	&	15	&	272	&		"		&		\\
2006 Jul 04	&	03:27	&	07:33	&	15	&	228	&	$\sim0.019$	&	D06	\\
\bottomrule
\end{tabular}
\end{table*}

\section{Ephemeris and Radial velocity curves}
\label{sec:ephemeris}
The analysis carried out in this section was completed for the sole purpose of determining a revised ephemeris in order to improve the quality of the Roche tomograms in Section~\ref{sec:surfacemaps}. New ephemerides were determined from the radial-velocity curves independently for each AE~Aqr data set, by cross-correlation with a spectral-type template star, following \cite{watson2006} and \cite{hill2014}. The details of the template star used for each data set are shown in Table~\ref{tab:templatestars}, where the systemic velocity was measured by a Gaussian fit to the least-squares deconvolution (LSD, see Section~\ref{sec:lsd}) line profile for each star (using a line-list where lines with a central depth shallower than 10~per~cent of the continuum were excluded).

\begin{table}
\centering
\begin{tabular}{llc}
\toprule
Data set	&	Star	&	$\gamma$ (\kms)	\\
\midrule					
D04	&	HD 214759	&	$1.9091\pm0.0014$	\\
D05, D06	&	HD 24916	&	$-5.144\pm0.005$	\\
D08	&	HD 187760	&	$-21.545\pm0.003$	\\
\bottomrule
\end{tabular}
\caption{Spectral-type templates used to calculate new ephemerides for AE Aqr. Columns 1-3 list the data for which the template star was used, the star's designation, and its systemic velocity.} 
\label{tab:templatestars}
\end{table}

\begin{table*}
\centering
\begin{tabular}{lcccc}
\toprule
Data	&	T$_{0}$ (HJD)					&	$\gamma$ (\kms)	&	\kb (\kms)	&	\vsini (\kms)	\\
\midrule													
D01	&	$	2452131.31345	\pm	0.00007	$	&	$-59.0\pm1.0$	&	$168.4\pm0.2$	&	-	\\
D04	&	$	2453195.440623	\pm	0.000013	$	&	$-60.065\pm0.024$	&	$169.82\pm0.03$	&	98.9	\\
D05	&	$	2453588.571407	\pm	0.000016	$	&	$-59.33\pm0.04$	&	$169.15\pm0.05$	&	99.2	\\
D06	&	$	2453921.60201	\pm	0.00005	$	&	$-59.53\pm0.18$	&	$165.85\pm0.21$	&	97.6	\\
D08	&	$	2454684.399364	\pm	0.000010	$	&	$-61.652\pm0.018$	&	$166.049\pm0.025$	&	102.8	\\
D09 \& D09b	&	$	2455071.356125	\pm	0.000009	$	&	$-62.072\pm0.017$	&	$167.433\pm0.022$	&	100.1	\\
\bottomrule
\end{tabular}
\caption{The new ephemerides for each data set of AE Aqr based on the radial velocity analysis described in Section~\ref{sec:ephemeris}. Columns 1-4 list the data set, the ephemeris and associated statistical uncertainty, as calculated from the fit to the RV curves, the systemic velocity, and the radial velocity semi-amplitude. Only the ephemerides are adopted for the Roche tomography analysis.} 
\label{tab:ephemeris}
\end{table*}

For this, we restricted ourselves to the spectral regions lying between $6000-6270$~\r{A} and $6320-6500$~\r{A}, as these contain strong absorption lines from the secondary star, and reduce the probability of introducing a continuum slope from the blue primary. Both the AE Aqr and K4V template spectra were normalized by dividing by a constant, and the continuum was fit using a third-order polynomial, and subtracted, thus preserving the line strength. The template spectrum was then artificially broadened (initially by 100~\kms) to account for the rotational velocity (\vsini) of the secondary, multiplied by a constant, and subtracted from an averaged high-signal-to-noise orbitally-corrected AE Aqr spectrum. These latter three steps were repeated, artificially broadening the template spectrum in 0.1~\kms steps until the scatter in the residual spectrum was minimized. This typically took two to three iterations. Through the above process, a cross-correlation function (CCF) was calculated for each AE Aqr spectrum, and the peak of the CCF was found using a parabolic fit. A radial velocity (RV) curve was then derived by fitting a sinusoid through the CCF peaks, obtaining new zero-point ephemerides for each data set (shown in Table~\ref{tab:ephemeris}), with the orbital period fixed at $P_{\text{orb}} = 0.41165553$~d (from \citealt{casares1996}). All subsequent analysis of each data set have been phased with respect to these new ephemerides. Separate ephemerides were calculated for each data set as the RV curves are affected by systematics, and so combining all data to calculate a single global ephemeris may not be optimal. Furthermore, the scatter in the O--C values of all data is relatively small, with a standard deviation of 0.12~per~cent of the orbital period.

The RV measurements obtained from the cross-correlation method described above are relatively insensitive to the use of a poorly-matched template, or an incorrect amount of template broadening. Surface features such as irradiation or star spots, as well as the tidal distortion of the secondary are more likely to introduce systematic errors in RV measurements, if not properly accounted for (e.g. \citealt{davey1992}). In addition, no detailed attempt was made to determine the best-fitting spectral-type or binary parameter determination in this analysis, however, for completeness we include the systemic velocity ($\gamma$), the radial velocity semi-amplitude (\kb) and the projected rotational velocity (\vsini) in Table~\ref{tab:ephemeris}.

\begin{figure*}
\centering
\begin{minipage}{0.5\textwidth}
\centering
\includegraphics[width=\textwidth]{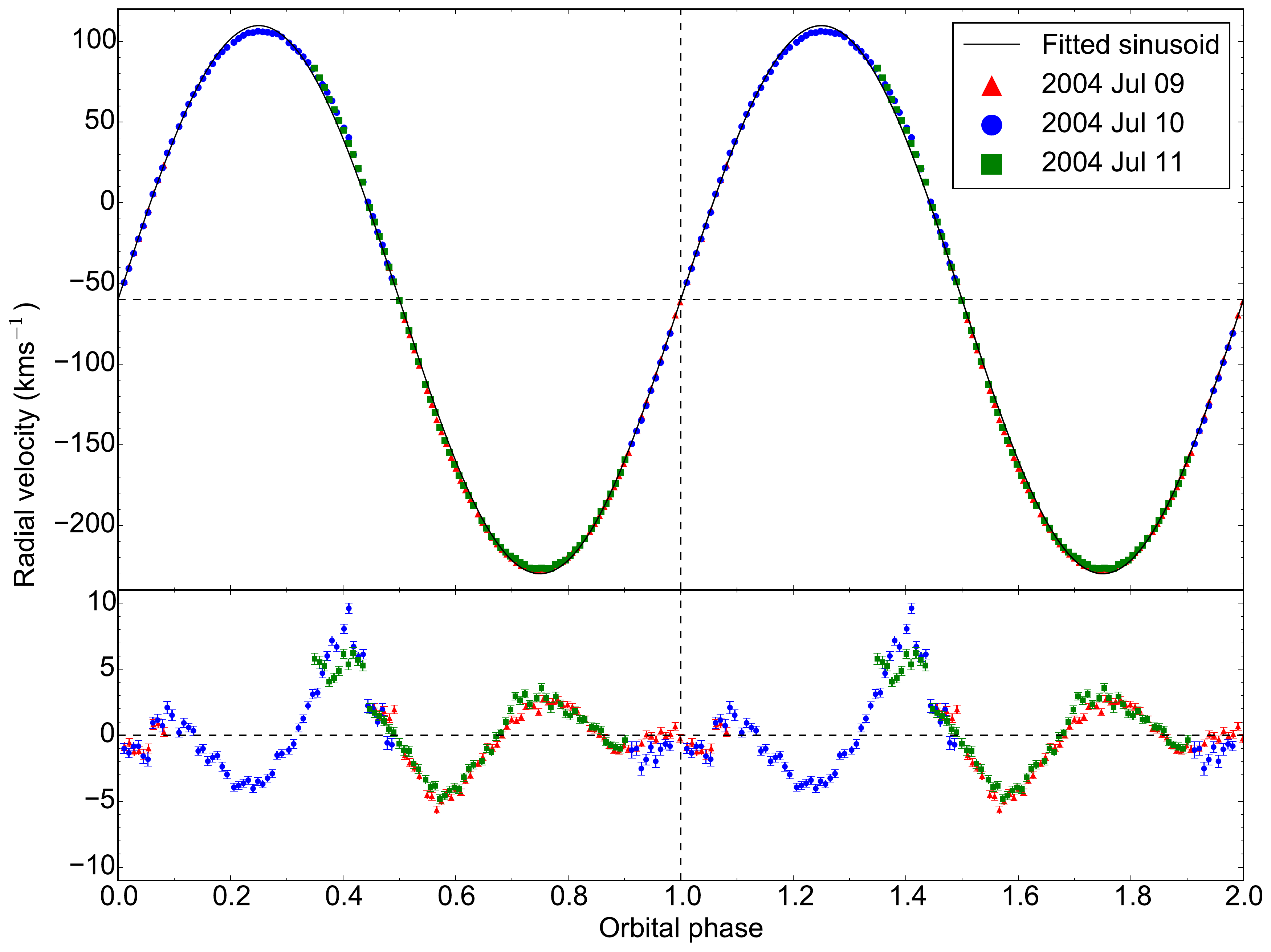}
\includegraphics[width=\textwidth]{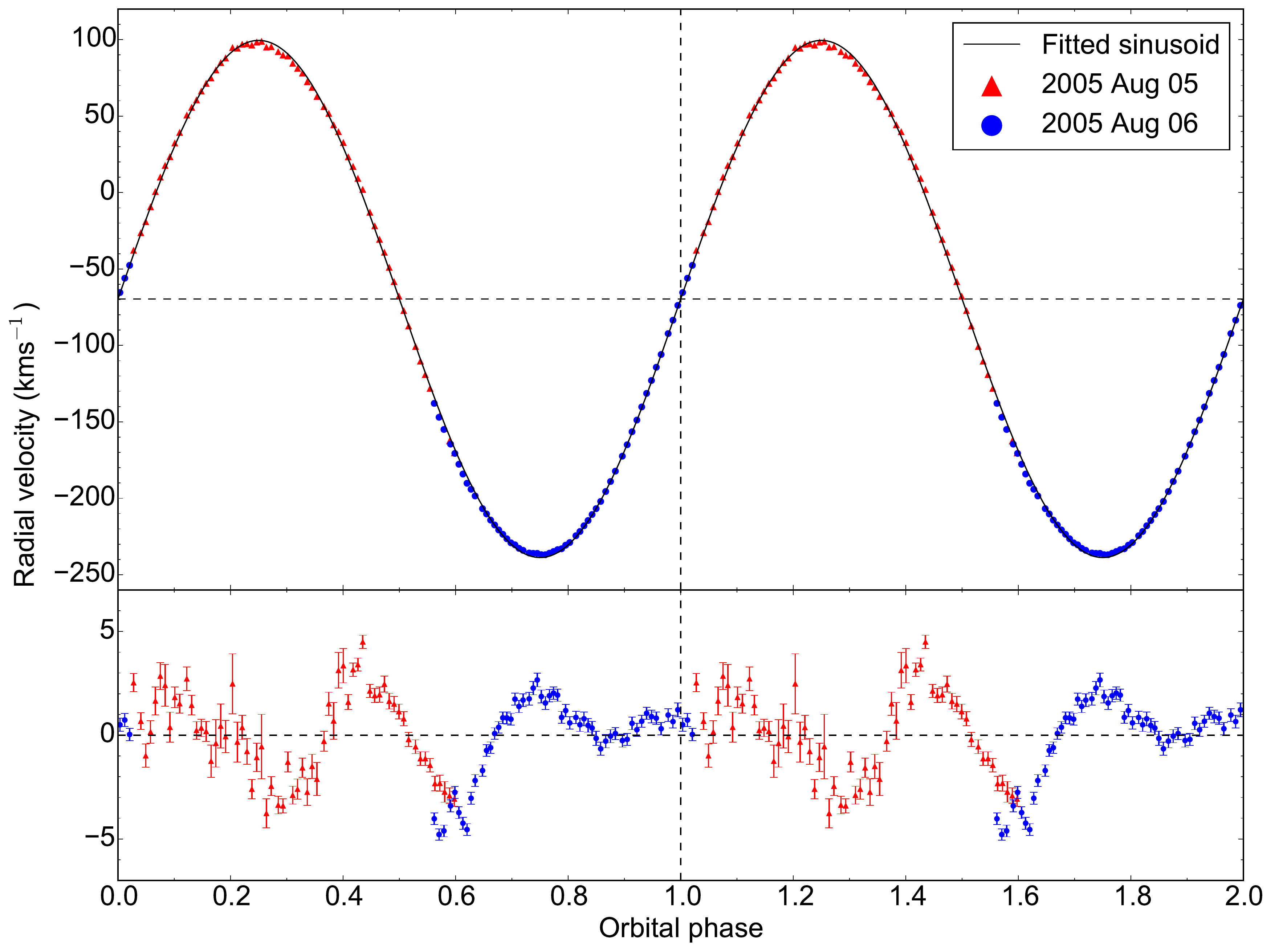}
\end{minipage}\hfill
\begin{minipage}{0.5\textwidth}
\centering
\includegraphics[width=\textwidth]{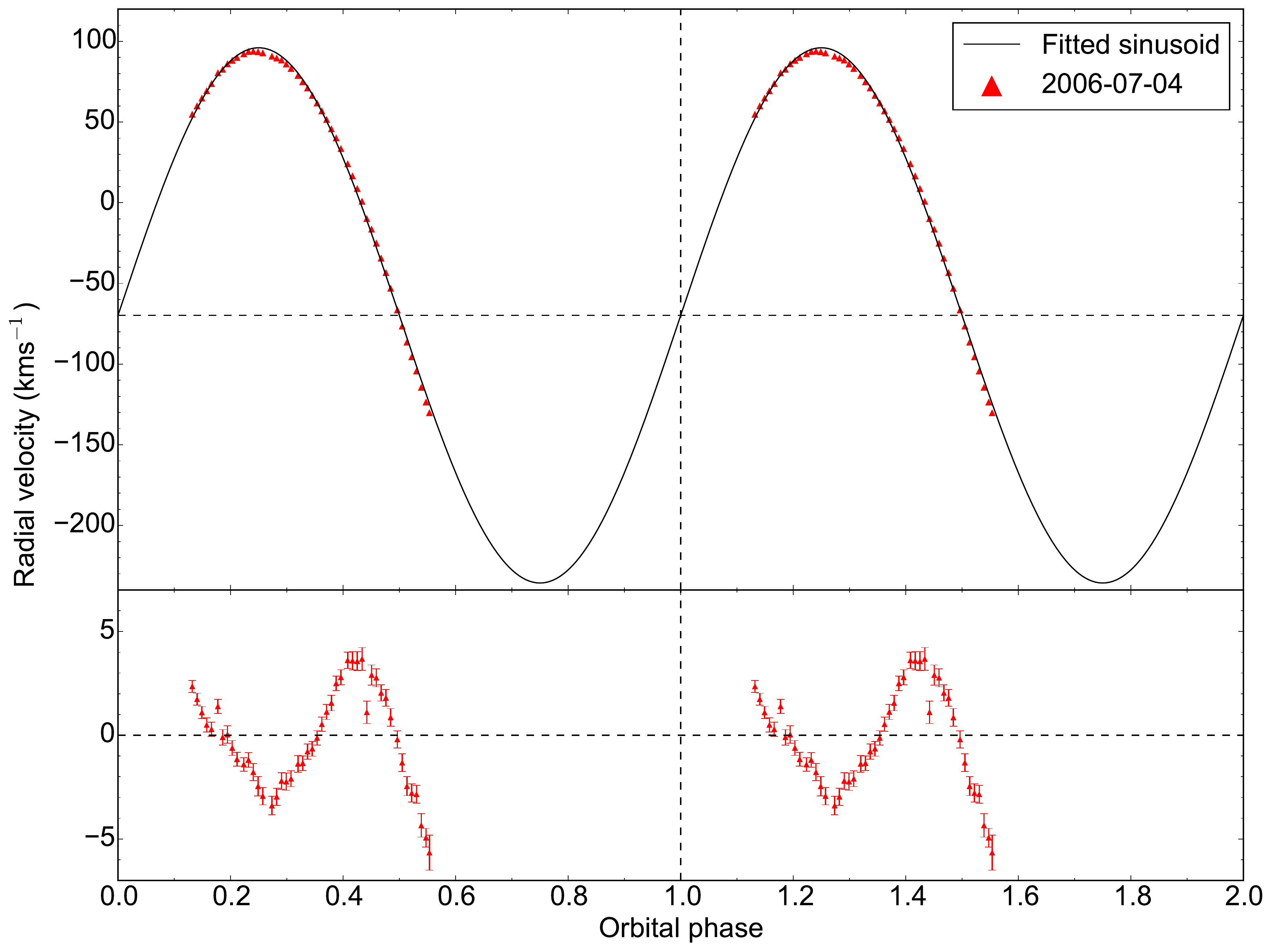}
\includegraphics[width=\textwidth]{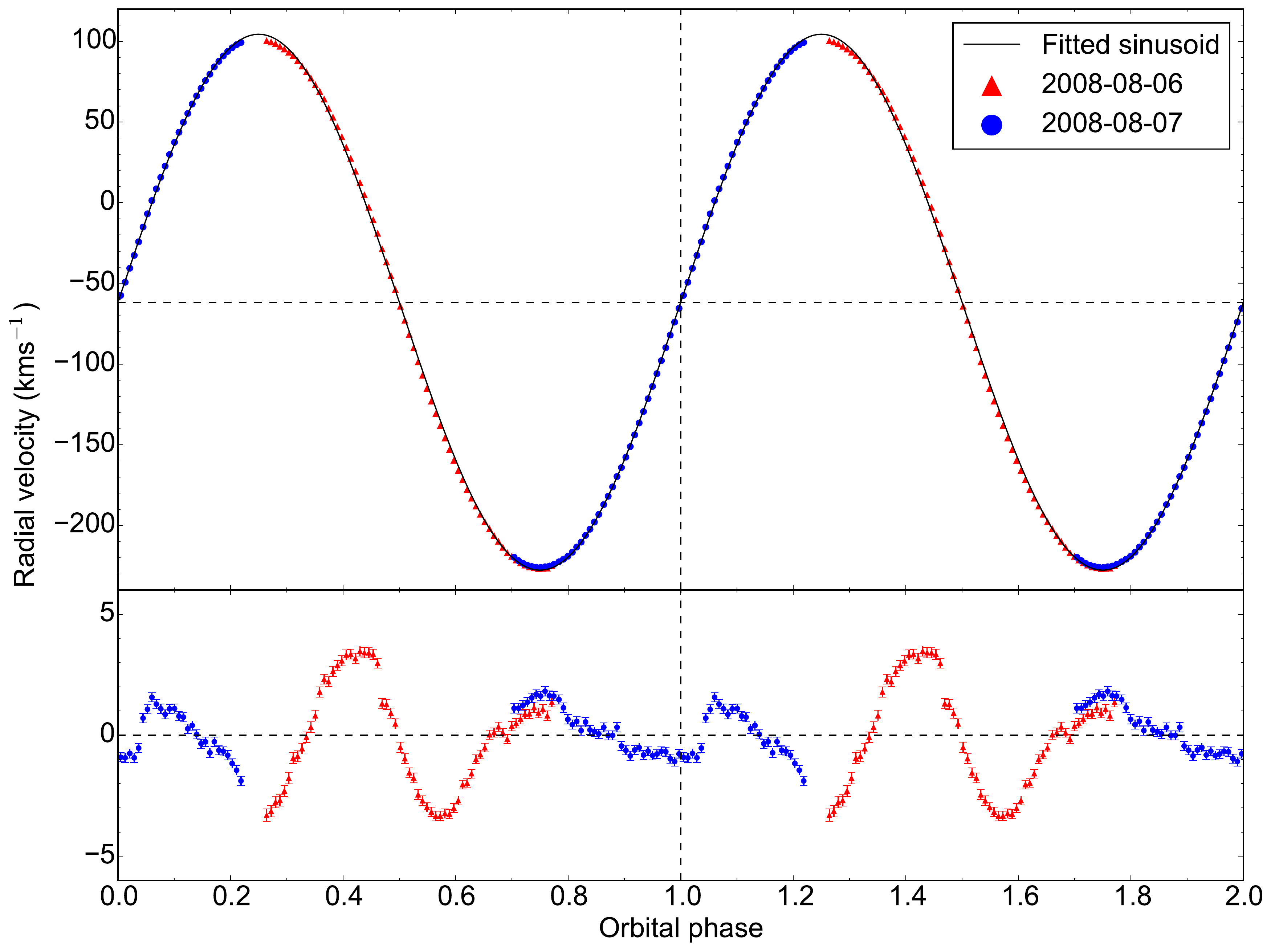}
\end{minipage}
\label{fig:rvcurve}
\caption{The radial velocity curves of AE Aqr for D04 (top left), D05 (bottom left), D06 (top right) and D08 (bottom right). The points are phase folded for clarity using the ephemerides in Table~\ref{tab:ephemeris}, and a least-squares sinusoid fit to the RV points (assuming a circular orbit) is shown as a solid line. The lower panels of each plot show the residuals after subtracting the fitted sinusoid, as well as the statistical uncertainies of the measured RVs.}
\end{figure*}

Figure~\ref{fig:rvcurve} shows the measured radial velocities for each data set, the fitted sinusoid, and the residuals after subtracting the sinusoid. The inherent systematic biases are clearly evident as deviations from a perfect sinusoid, and as such, the binary parameters derived from this RV analysis have not been used in the subsequent analysis presented in this work.

The small variation in $\gamma$, \kb and \vsini between data sets (see Table~\ref{tab:ephemeris}) may be due to instrumental offsets between observations, but the spread in values is most likely dominated by the systematic biasing of RV measurements due to surface features such as irradiation and starspots. Such features may alter the slope of the RV curve, changing \kb, and due to their non-uniform distribution, surface features may cause a shift in the measured $\gamma$. We note that the $1\sigma$ spread in ephemerides is~$\sim37$~s, suggesting the period was stable over the duration of all observations.

\begin{table}
\centering
\caption{Spectral regions excluded from analysis.} 
\label{tab:wavemasked}
\begin{tabular}{cl}
\toprule
Masked region (\r{A})	&	Comments	\\
\midrule			
$<4600$	&	Noisy, He~\textsc{i} \& H emission	\\
$4670 - 4700$	&	He~\textsc{ii} emission	\\
$4830 - 4885$	&	H$\beta$ emission	\\
$5850 - 5900$	&	He~\textsc{i} emission \& Na~\textsc{i} doublet	\\
$6270 - 6320$	&	Tellurics	\\
$6525 - 6600$	&	H$\alpha$ emission	\\
$6650 - 6700$	&	He~\textsc{i} emission	\\
$6864 - 6970$	&	Tellurics	\\
$7031 - 7093$	&	He~\textsc{i} emission	\\
$7158 - 7370$	&	Tellurics	\\
$7590 - 7705$	&	Tellurics	\\
\bottomrule
\end{tabular}
\end{table}

\section{Roche tomography}
\label{sec:rochetomography}
Roche tomography is a technique analogous to Doppler imaging (e.g. \citealt{vogt1983}), and is specifically designed to indirectly image the secondary stars in close binaries such as CVs (\citealt{rutten1994,rutten1996,schwope2004,watson2003,watson2006,watson2007,dunford2012,hill2014}), pre-CVs (e.g.~\citealt{parsons2015}) and X-ray binaries (e.g.~\citealt{shahbaz2014}). The technique assumes that the secondary is locked in synchronous rotation with a circularized orbit, and that the star is Roche-lobe filling. We refer the reader to the references above and the technical reviews of Roche tomography by \cite{watson2001} and \cite{dhillon2001} for a detailed description of the axioms and methodology.

\section{Least squares deconvolution}
\label{sec:lsd}
Least squares deconvolution (LSD) was applied to all spectra in the same manner as in \cite{watson2006}, producing mean line profiles with a substantially increased signal-to-noise ratio (SNR). LSD requires that the spectral continuum be flattened. However, the contribution to each spectrum from the accretion regions is unknown, and a constantly changing continuum slope due to, for example, flaring (see Figure~\ref{fig:photometry}) or the varying aspect of the accretion regions, means a master continuum fit to the data cannot be used. In addition, as the contribution of the secondary star to the total light of the system is constantly varying, normalizing the continuum by division would result in the photospheric absorption lines from the secondary star varying in relative strength from one exposure to the next. Hence, we are forced to subtract the continuum from each spectrum. This was achieved by fitting a spline to the data. As the spectral type of AE Aqr has been determined to lie in the range K3-K5V (\citealt{crawford1956,chincarini1981,tanzi1981,bruch1991}), we generated a stellar line list for a K4V type star ($T_{\mathrm{eff}} = 4750$~K and $\log{g} = 4.5$, the closest approximation available) using the Vienna Atomic Line Database (VALD, see \citealt{kupka2000}), adopting a detection limit of 0.2. The normalized line depths were scaled by a fit to the continuum of a K4V template star so each line's relative depth was correct for use with the continuum subtracted spectra.

Emission lines and telluric lines were masked in the spectra and line list -- the excluded spectral regions are detailed in Table~\ref{tab:wavemasked}. This meant that over the 4600--7700~\r{A} spectral range for D04, D05 and D06, 2354 lines were available over which to carry out LSD. Similarly, 1558 lines were available for D08 in the spectral range 4780--6810~\r{A}. After carrying out LSD, a small continuum slope was present in the LSD profiles. This was removed by masking out the line centre and subtracting a second-order polynomial which was fit to the continuum. Details for D01 and D09a~\&~D09b may be found in \cite{watson2001} and \cite{hill2014}, respectively.

The variable light contribution of the secondary means we cannot normalize the data in the usual way. Instead, we are forced to use relative line fluxes, requiring the spectra to be slit-loss corrected. For D04, D05 and D06 we used simultaneous photometry to monitor transparency and target brightness variations (see Section~\ref{sec:photometry}). We corrected for slit losses by dividing each LSD profile by the ratio of the flux in the spectrum (after integrating the spectrum over the photometric filter response function) to the corresponding photometric flux. The value of photometric flux used was the mean over the duration of the spectroscopic exposure. As we were unable to obtain simultaneous photometry for D08, we used the fits from Roche tomography to iteratively scale the LSD line profiles in the same manner as carried out for D09a~\&~D09b in~\cite{hill2014}. The resulting scaled profiles were visually inspected and found to be consistent.

\begin{figure*}
\centering
\includegraphics[width=0.6\textwidth,angle=270]{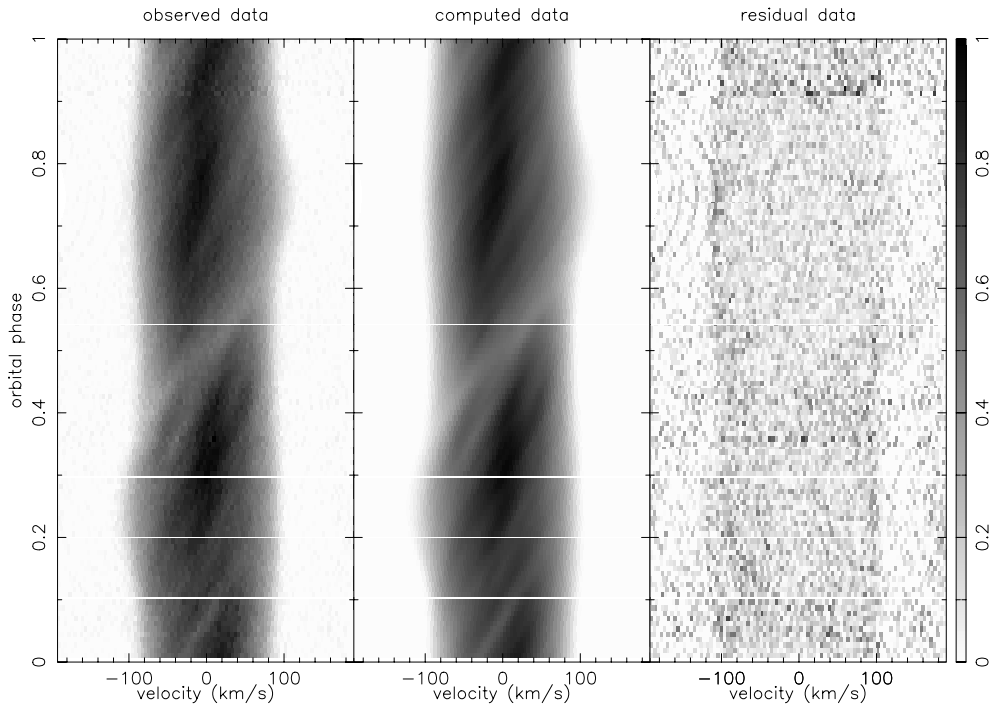}
\caption{Trailed LSD profiles of AE Aqr for D04. The orbital motion has been removed assuming the binary parameters found in Section~\ref{sec:incsystemicmasses}, allowing individual starspot tracks across the profiles and the variation in \vsini to be more clearly observed. Panels show (from left to right) the observed LSD profiles, the computed fits to the data using Roche tomography, and the residuals (increased by a factor of 10). Starspots and surface features appear bright in these panels, where a grey-scale of 1 corresponds to the maximum line depth in the reconstructed profiles. }
\label{fig:trails04}
\end{figure*}

\begin{figure*}
\centering
\includegraphics[width=0.6\textwidth,angle=270]{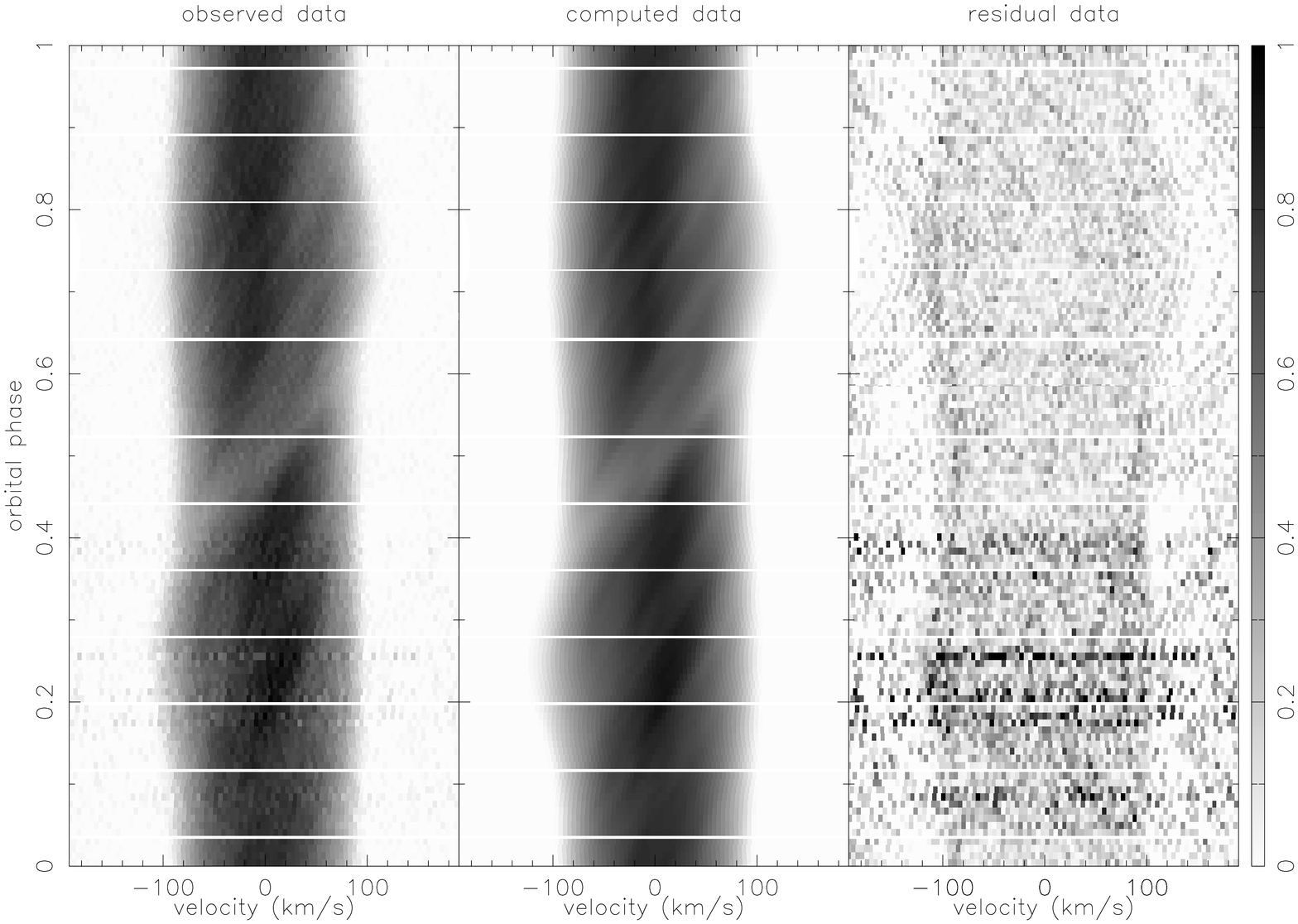}
\caption{The same as Figure~\ref{fig:trails04} but for D05.}
\label{fig:trails05}
\end{figure*}

\begin{figure*}
\centering
\includegraphics[width=0.6\textwidth,angle=270]{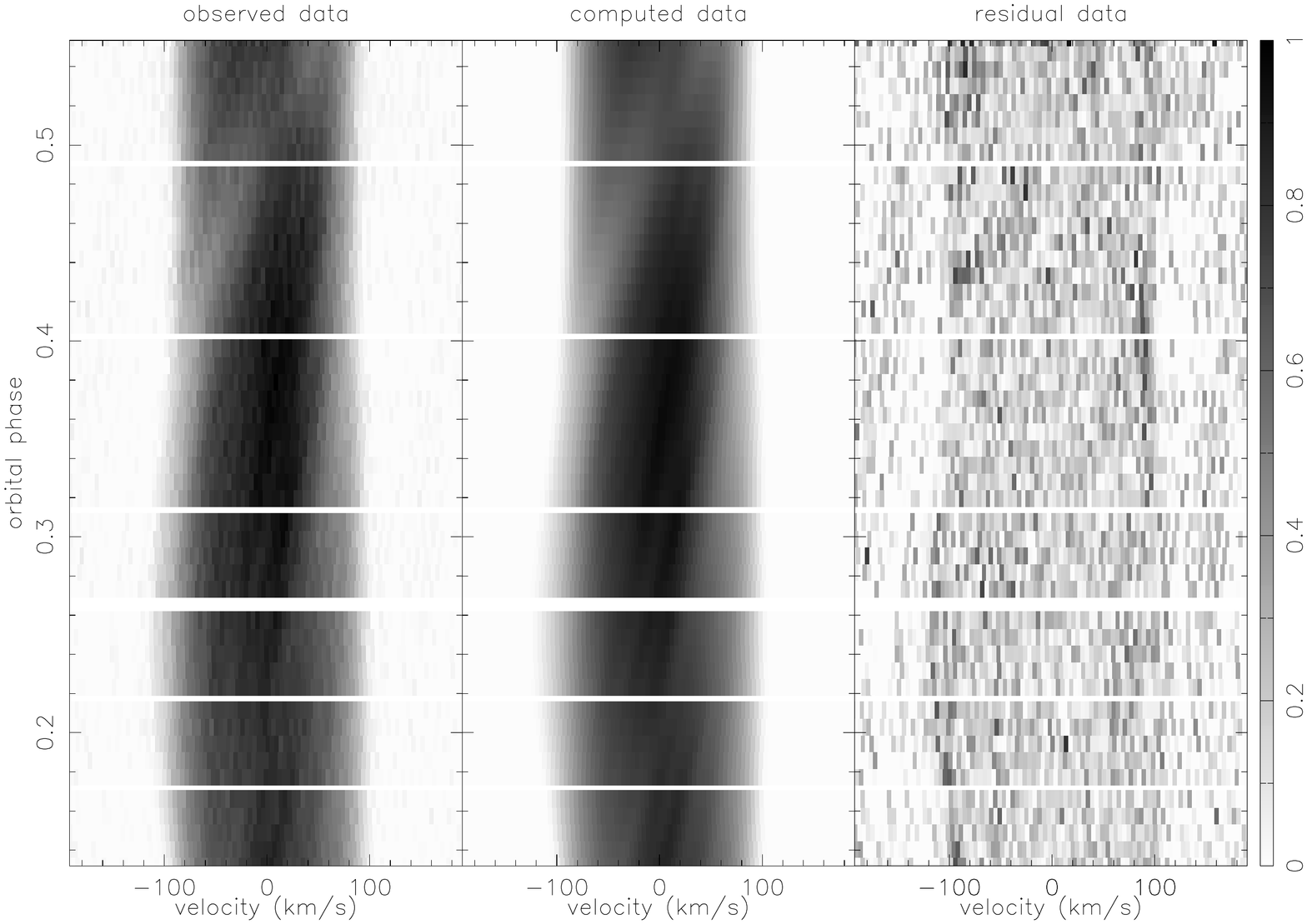}
\caption{The same as Figure~\ref{fig:trails04} but for D06.}
\label{fig:trails06}
\end{figure*}

\begin{figure*}
\centering
\includegraphics[width=0.6\textwidth,angle=270]{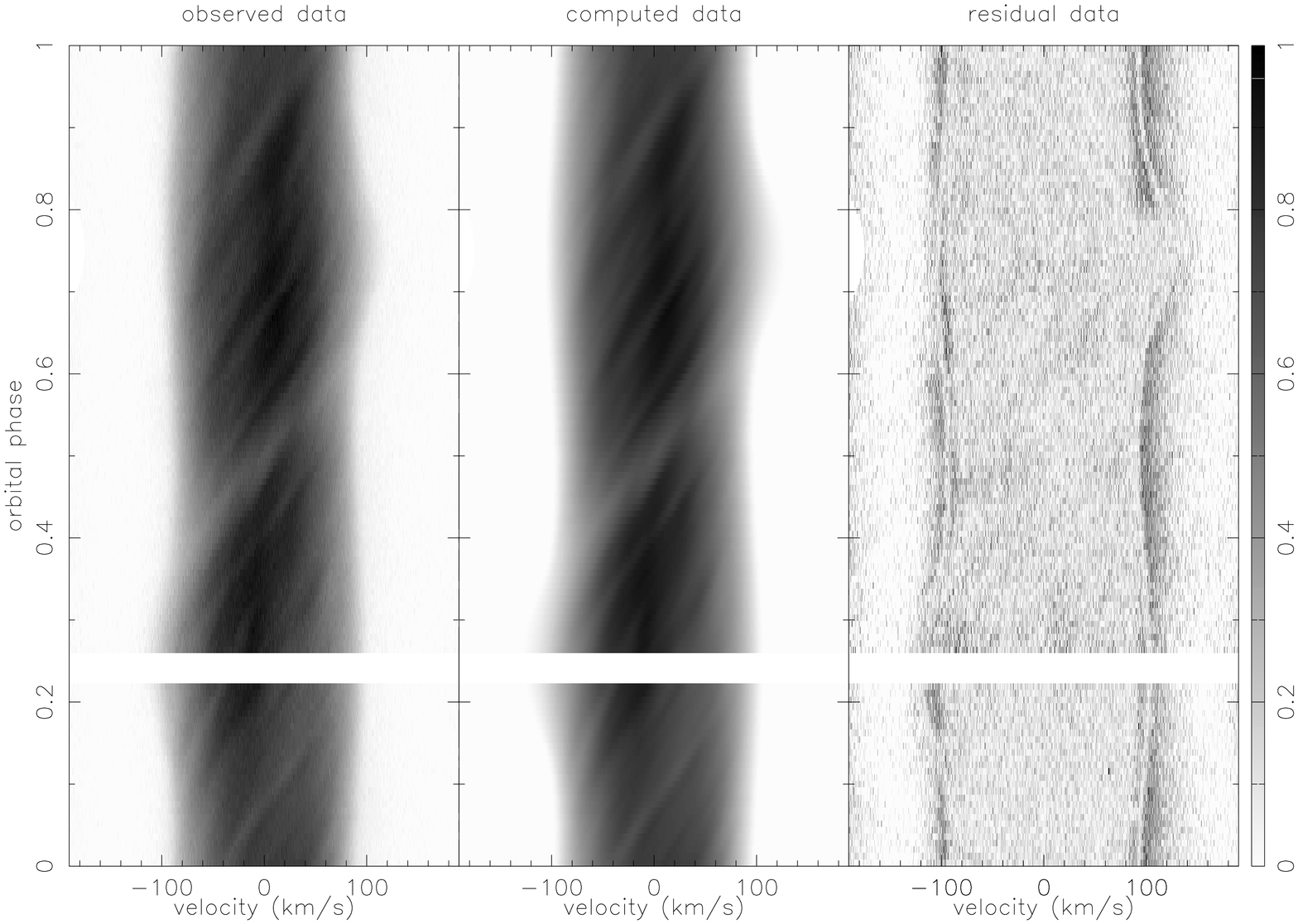}
\caption{The same as Figure~\ref{fig:trails04} but for D08. The fits to the wings of the LSD profiles appear relatively poor, but due to the higher SNR of the data, the absolute level of the residuals is actually lower (see text for discussion).}
\label{fig:trails08}
\end{figure*}

The final LSD profiles, the computed fits, and the residuals after subtracting the fits from the LSD profiles, are trailed for each data set in Figures~\ref{fig:trails04}~to~\ref{fig:trails08}, where the orbital motion has been removed. Starspots and surface features are clearly visible as emission bumps moving through the profiles from negative to positive velocities as AE Aqr rotates. The variation in \vsini due to the tidal-distortion is also clearly apparent. Furthermore, the residuals of D04, D05 and D08 (see Figures~\ref{fig:trails04},~\ref{fig:trails05}~and~\ref{fig:trails08}) show narrow emission features that are seen to lie outside the stellar absorption profile, and appear to move in anti-phase with respect to the secondary. Similar features were also seen in the trails of D09a~\&~D09b (see Figure~3 of \citealt{hill2014}), and as they are visible at all phases, they may be due to accretion material in the system. However, as this emission is weak, we did not carry out any further analysis. The residuals also show the relatively poor fit to the wings of the LSD profiles, resulting from adopting a spherical limb darkening law for a non-spherical object. We note that the SNR of D08 is significantly higher than for any other data set (see Table~\ref{tab:obs}), resulting in a relatively better fit with Roche tomography. This means that visually, the fits to the wings of the LSD profiles appear of lower quality, but the absolute level of the residuals is indeed lower than that of the other data sets.

\section{System parameters}
\label{sec:systempars}
\begin{figure}
\centering
\includegraphics[width=0.4\textwidth]{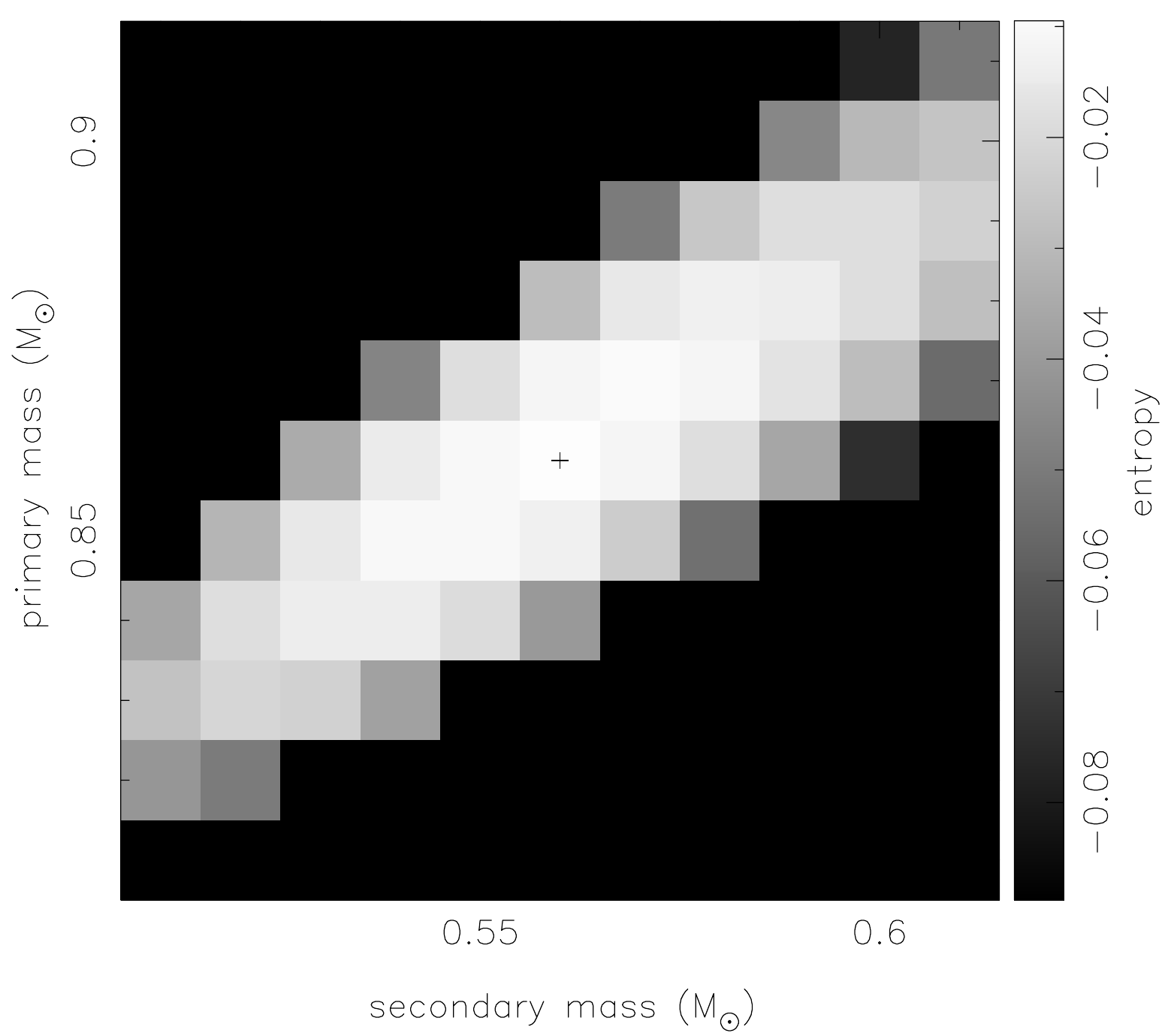}
\caption{The entropy landscape for AE Aqr using D05, assuming the parameters given in Table~\ref{tab:systemparameters}. Dark regions indicate masses for which no acceptable solution could be found. The cross marks the point of maximum entropy, corresponding to component masses of $M_{1} = 0.86~\mathrm{M}_{\sun}$ and $M_{2} = 0.56~\mathrm{M}_{\sun}$.}
\label{fig:entland}
\end{figure}

The system parameters (systemic velocity $\gamma$, orbital inclination $i$, primary star mass $M_{1}$ and secondary star mass $M_{2}$) of AE Aqr were determined using the standard methodology of Roche tomography (e.g. \citealt{watson2006,watson2007}). Adopting incorrect system parameters when carrying out Roche tomography reconstructions results in spurious artefacts in the final image. These artefacts are well characterised (see \citealt{watson2001}), and always increase the amount of structure (information content) of the map, decreasing the map entropy. We can constrain the binary parameters by carrying out map reconstructions for many pairs of component masses, fitting to the same $\chi^{2}$. This can be visualized as an entropy landscape, with an example shown in Figure~\ref{fig:entland}, where the optimal masses are the pair that produce the map of maximum entropy (least information content). Entropy landscapes are then repeated for different values of $i$ and $\gamma$, with the optimal set of parameters those which produce the map containing least structure (the map of maximum entropy). 

The optimal system parameters are unique to each data set, as systematic effects may result in different optimal parameters between data sets. Hence, we do not adopt the mean values across all data sets for our analysis, as to do so may increase the number of artefacts reconstructed in the maps. This is further discussed in Section~\ref{sec:surfacemaps}.

\subsection{Limb Darkening}
\label{sec:limbdarkening}
Following \cite{hill2014}, we adopted the four-parameter non-linear limb darkening model of \cite{claret2000}. The stellar parameters closest to that of a K4V star were adopted, which for the PHOENIX model atmosphere were $\log{g} = 4.5$ and $T_{\mathrm{eff}} = 4800$~K. The adopted coefficients for each data set are shown in Table~\ref{tab:limbcoeff}, where different (but very similar) values were used for each data set due to the different central wavelengths of the spectra. The treatment of limb darkening for D01 and D09a \& D09b may be found in the previously published work of \cite{watson2006} and \cite{hill2014}, respectively.

\begin{table}
\centering
\caption{Limb-darkening coefficients.}
\label{tab:limbcoeff}
\begin{tabular}{crrrr}
\toprule
Coefficient & D04 & D05 & D06 & D08 \\
\midrule
$a_{1}$	&	0.724	&	0.724	&	0.724	&	0.724	\\
$a_{2}$	&	-0.759	&	-0.768	&	-0.764	&	-0.764	\\
$a_{3}$	&	1.349	&	1.356	&	1.353	&	1.353	\\
$a_{4}$	&	-0.415	&	-0.411	&	-0.412	&	-0.412	\\
\bottomrule
\end{tabular}
\end{table}

\subsection{Systemic velocity, inclination and masses}
\label{sec:incsystemicmasses}
All data sets were fit independently. For each, we constructed a series of entropy landscapes for a range of orbital inclinations $i$ and systemic velocities $\gamma$. For given values of $i$ and $\gamma$ we selected the pair of masses that produced the map of maximum entropy. The results of this analysis are presented here.

\subsubsection{Systemic velocity}
\label{sec:systemic}
\begin{figure}
\centering
\includegraphics[width=0.5\textwidth]{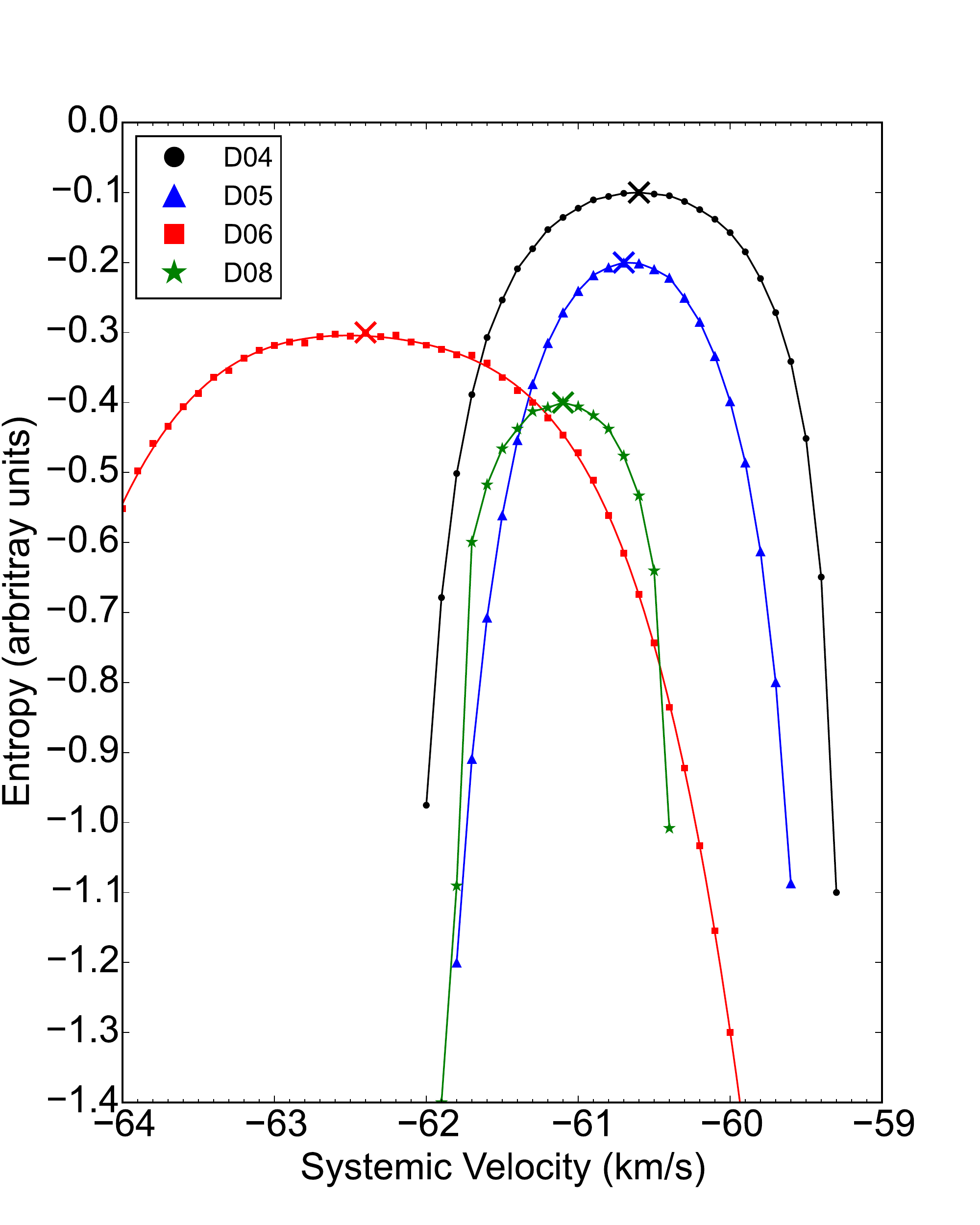}
\caption{The points show the maximum entropy obtained
in each data set as a function of systemic velocity for AE Aqr. The optimal inclination and masses were adopted for each data set, as found in Sections~\ref{sec:inclination}~\&~\ref{sec:masses}, respectively. The points are offset in the ordinate for clarity. Crosses mark the optimal value of $\gamma$, and solid lines are shown only as a visual aid.}
\label{fig:systemic}
\end{figure}

Figure~\ref{fig:systemic} shows the map entropy (after adopting optimal values of $M_{1}, M_{2}~\&~i$) as a function of systemic velocity, for each data set. Crosses mark the peak of the `entropy parabola', giving the optimal values of $\gamma$, as listed in Table~\ref{tab:systemparameters}. The measured values of $\gamma$ are consistent with that of previous work, falling within the uncertainties given by both \cite{welsh1995} and \cite{casares1996}. The significant difference between the $\gamma$ found here and that found by \cite{echevarria2008} stems from the uncertainties in the absolute radial velocities of the template stars used in the latter authors' analysis. The spread in $\gamma$ as measured by Roche tomography may be explained by instrumental offsets between different instruments, as well as for the same instrument over different observation periods. Values of $\gamma$ obtained by using entropy landscapes was found to be independent of the assumed inclination, as previously found by \cite{watson2003,watson2006} and \cite{hill2014}. In addition, the values obtained using the radial velocity curves are similar, although these will be biased, as discussed in section~\ref{sec:ephemeris}. 

\subsubsection{Inclination}
\label{sec:inclination}
\begin{figure}
\centering
\includegraphics[width=0.5\textwidth]{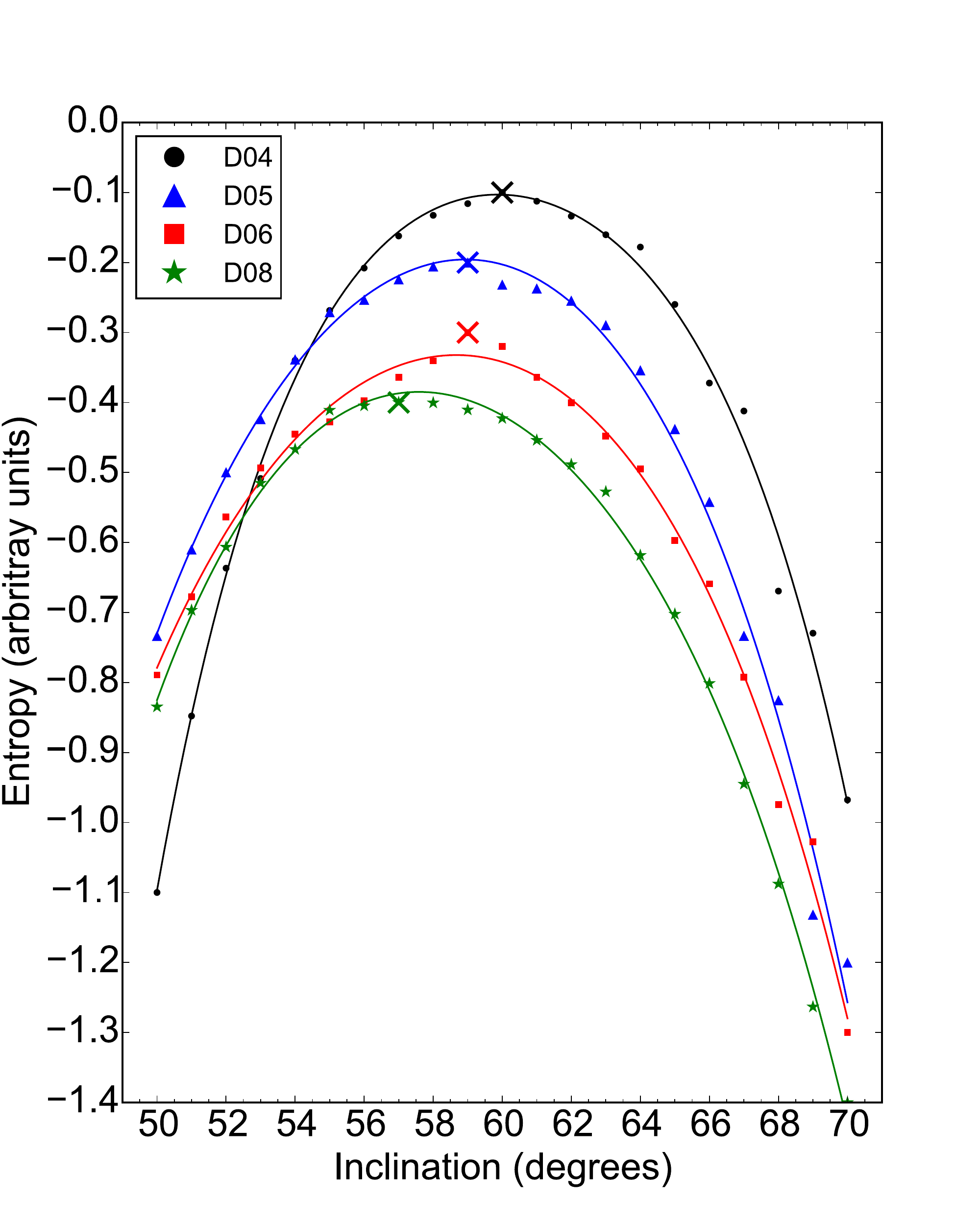}
\caption{The points show the maximum entropy obtained for each data set as a function of inclination, assuming the optimal values of $\gamma$ and masses, as found in Sections~\ref{sec:systemic}~\&~\ref{sec:masses}, respectively. Crosses mark the optimal value of $i$, and a solid line is shown as a visual aid.}
\label{fig:inclination}
\end{figure}

Figure~\ref{fig:inclination} shows the maximum entropy obtained as a function of inclination, for each data set, assuming the systemic velocities derived in Section~\ref{sec:systemic}. Crosses mark the maximum entropy obtained for a given data set, and the corresponding inclinations are listed in Table~\ref{tab:systemparameters}. All values of $i$ are consistent with previously published work by \cite{welsh1995} and \cite{casares1996}, although all values lie below that found by \cite{echevarria2008}. Furthermore, all values of $i$ lie between the previously determined inclinations of D01 ($i = 66\degr$, \citealt{watson2006}) and D09a~\&~D09b ($i = $50--51$\degr$, \citealt{hill2014}). 

The consistency of $i =$~57--60$\degr$ across the four data sets is a reassuring result, as inclination is the worst constrained parameter when using Roche tomography. We do not have a clear explanation for the discrepancy between the inclinations found here and that of D01 and D09a~\&~D09b.

\subsubsection{Masses}
\label{sec:masses}
The component masses of AE Aqr were determined using entropy landscapes, with an example shown for D05 in Figure~\ref{fig:entland}. For each data set we assumed $i$ and $\gamma$, as derived in Sections~\ref{sec:systemic}~\&~\ref{sec:inclination}. Our derived masses shown in Table~\ref{tab:systemparameters} are consistent (within the uncertainties) of those found by \cite{echevarria2008}, \cite{welsh1995}, and \cite{casares1996}, once the masses have been adjusted to account for the change in inclination. The differences between the masses determined here, and those found in previous studies of AE Aqr using Roche tomography, are simply due to the use of slightly different inclination values. Indeed the mass ratios $q$ are in excellent agreement with previous work, and are typically 6~per~cent larger than those found by \cite{echevarria2008}, \cite{welsh1995}, and \cite{casares1996}. We note that the masses determined in this work are (in principle) the most reliable, as we correct for the systematic effects of surface features that may bias the measured RVs used to determine the system parameters in other work.  

\begin{figure}
\centering
\includegraphics[width=0.5\textwidth]{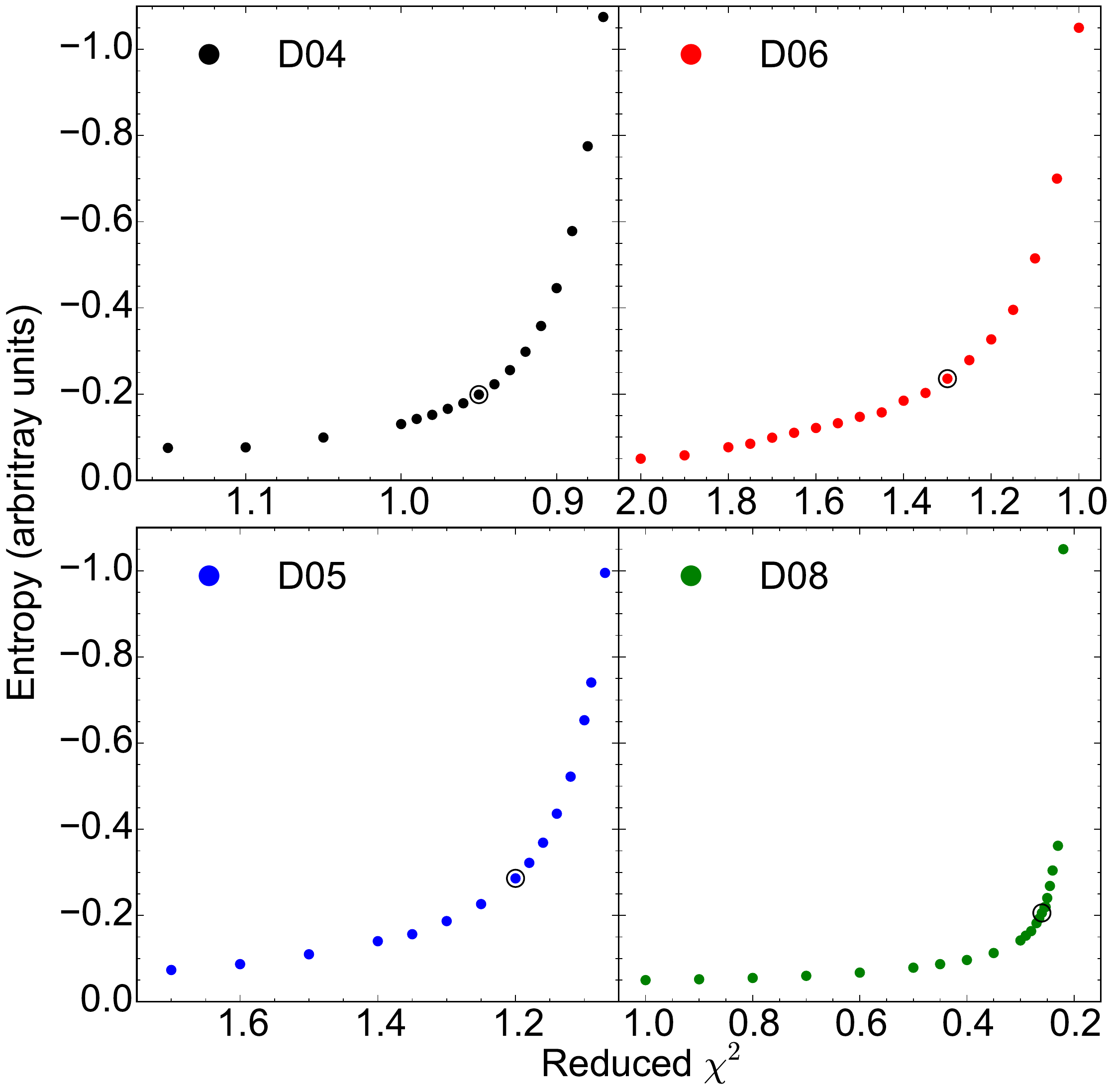}
\caption{The reconstructed-map entropy as a function of reduced $\chi^{2}$ for the Roche tomograms of each data set (marked in top left of panels). The system parameters derived in Section~\ref{sec:incsystemicmasses} were adopted for the fits. The selected aim $\chi^{2}$ is circled in each plot, and is taken as the point where there is a dramatic decrease in map entropy, and a corresponding increase in small scale features in the reconstruction. The final reduced~$\chi^{2}$ for D04, D05, D06 and D08 are 0.95,~1.2,~1.3~\&~0.26, respectively, where a $\chi^{2}<1$ indicates that we overestimated the size of our propagated uncertainties. See Section~\ref{sec:masses} for further discussion.}
\label{fig:chivsentropy}
\end{figure}

The target reduced~$\chi^{2}$ to which our data were fit was chosen as the point where the entropy of the reconstructed maps dramatically decreased when fits to a lower $\chi^{2}$ were performed. An increase in small scale structure contributes to a dramatic decrease in entropy, indicative of mapping noise in the Roche tomograms. Fitting to a higher reduced-$\chi^{2}$ caused fewer features to be mapped, and thus the system parameters were less well defined as more map pixels were assigned the default map value. Figure~\ref{fig:chivsentropy} shows how the map entropy depended on the aim $\chi^{2}$, where the adopted value is circled. The absolute value of $\chi^{2}$ is not a good reflection of the quality of fit, as a value above or below 1 indicates our error bars were systematically under or over estimated.

Assigning uncertainties to any of the derived system parameters ($i, \gamma, M_{1}, M_{2}$) is not trivial. As previously discussed in \cite{watson2001} and \cite{watson2006}, it would require using a Monte Carlo style technique combined with bootstrap resampling to generate synthetic data sets drawn from the same parent population as the observed data. Then, the same analysis carried out in this work would need to be applied to the hundreds of bootstrapped data sets, requiring an unfeasible amount of computation. Hence we do not assign strict uncertainties to our derived system parameters.

\begin{table*}
\caption{System parameters. Columns 1-6 list the data set or paper from which the parameters were taken, the systemic velocity as measured by Roche tomography, the inclination, the mass of the primary star, the mass of the secondary star and the mass ratio. The significantly higher component masses found in D09a~\&~D09b are due to the lower inclination found in that study.}
\label{tab:systemparameters}
\begin{tabular}{cccccc}
\toprule
Author / Data	&	$\gamma$ (\kms) 	&	$i$ (degrees) 	&	 $M_{1}$ (M$_{\sun}$) 	&	 $M_{2}$ (M$_{\sun}$) 	&	$q = \sfrac{\mathrm{M}_{2}}{\mathrm{M}_{1}}$	\\
\midrule											
D01	&	$-63$	&	66	&	0.74	&	0.50	&	0.68	\\
D04	&	$-60.6$	&	60	&	0.84	&	0.55	&	0.65	\\
D05	&	$-60.7$	&	59	&	0.86	&	0.56	&	0.65	\\
D06	&	$-62.4$	&	59	&	0.87	&	0.56	&	0.64	\\
D08	&	$-61.4$	&	57	&	0.94	&	0.64	&	0.68	\\
D09a	&	$-64.7\pm 2.1$	&	50	&	1.20	&	0.81	&	0.68	\\
D09b	&	$-62.9\pm 1.0$	&	51	&	1.17	&	0.78	&	0.67	\\
\cite{echevarria2008}	&	-63	&	$70\pm3$	&	$0.63\pm0.05$	&	$0.37\pm0.04$	&	0.60	\\
\cite{casares1996}	&	$-60.9\pm 2.4$	&	$58\pm6$	&	$0.79\pm0.16$	&	$0.50\pm0.10$	&	0.63	\\
\cite{welsh1995}	&	$-63\pm 3$	&	$54.9\pm7.2$	&	$0.89\pm0.23$	&	$0.57\pm0.15$	&	0.64	\\
\bottomrule
\end{tabular}
\end{table*}

\section{Surface maps}
\label{sec:surfacemaps}
\begin{figure}
\centering
\includegraphics[width=0.5\textwidth]{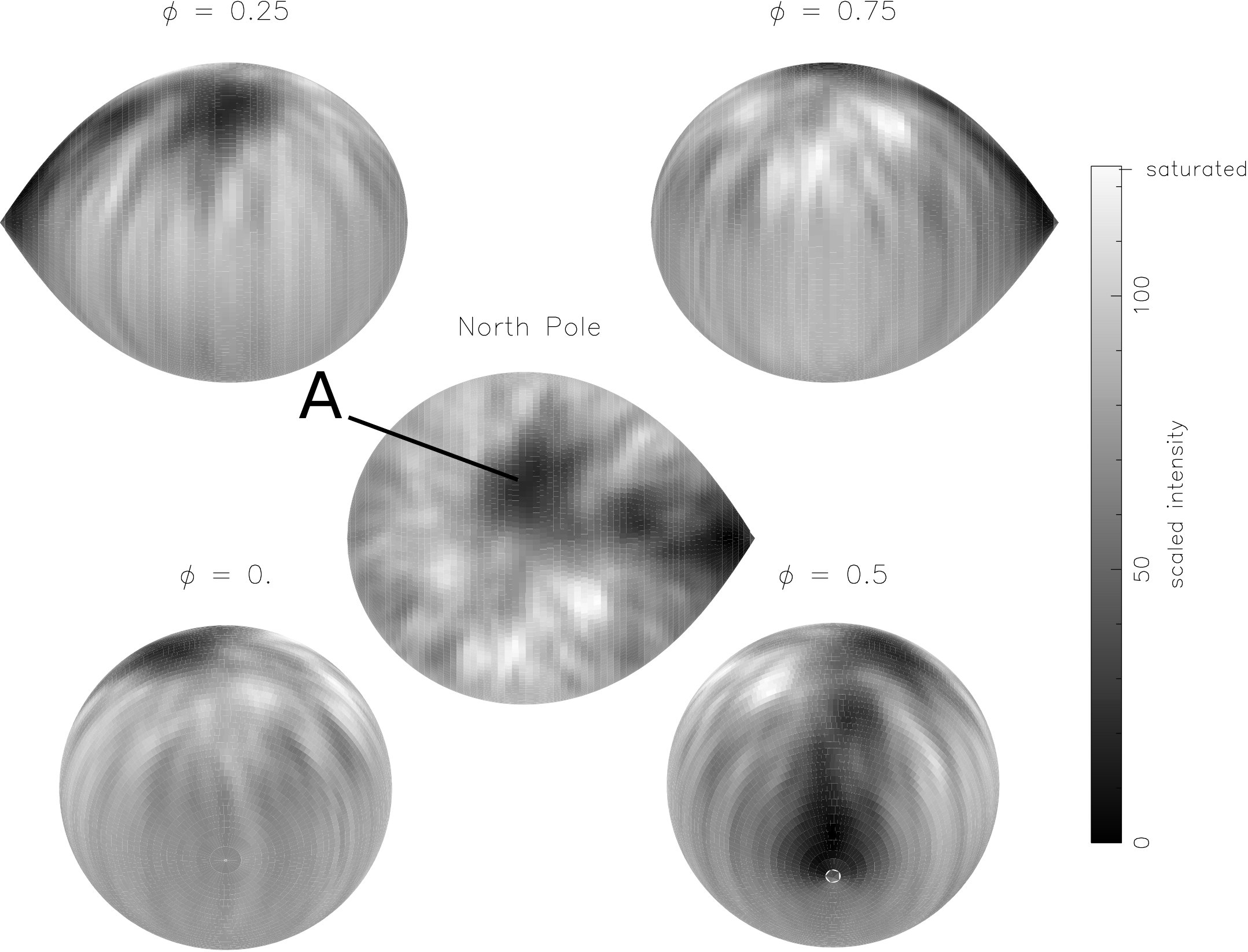}
\caption{The Roche tomogram of AE Aqr using D01. Dark grey-scales indicate regions of reduced absorption-line strength that is due to either the presence of starspots, the impact of irradiation, or gravity darkening. The absolute grey-scales are relative and are not necessarily comparable between maps. The orbital phase is indicated above each panel. Roche tomograms are shown without limb darkening for clarity.}
\label{fig:map01}
\end{figure}

\begin{figure}
\centering
\includegraphics[width=0.5\textwidth]{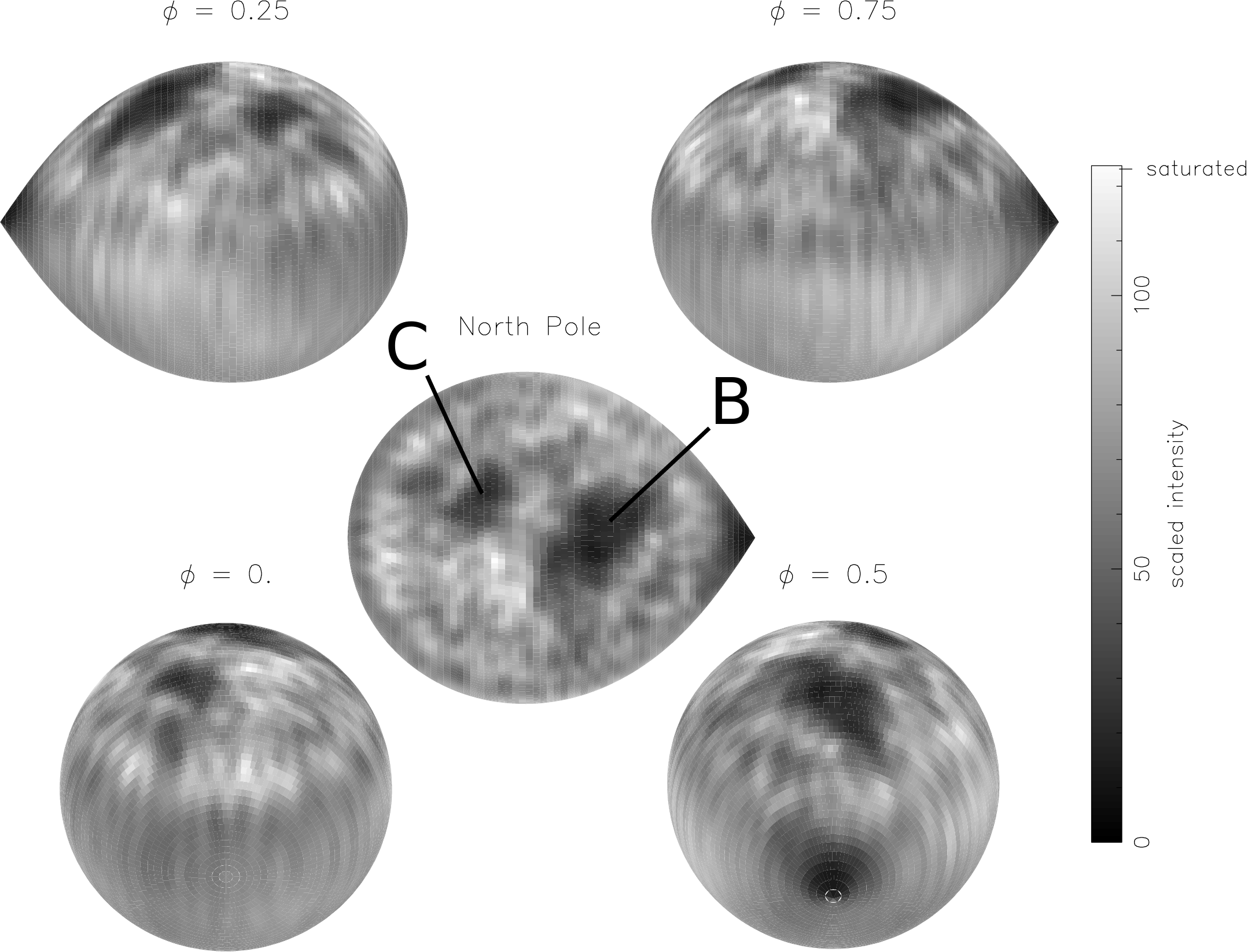}
\caption{The same as Figure~\ref{fig:map01} but for D04.}
\label{fig:map04}
\end{figure}

\begin{figure}
\centering
\includegraphics[width=0.5\textwidth]{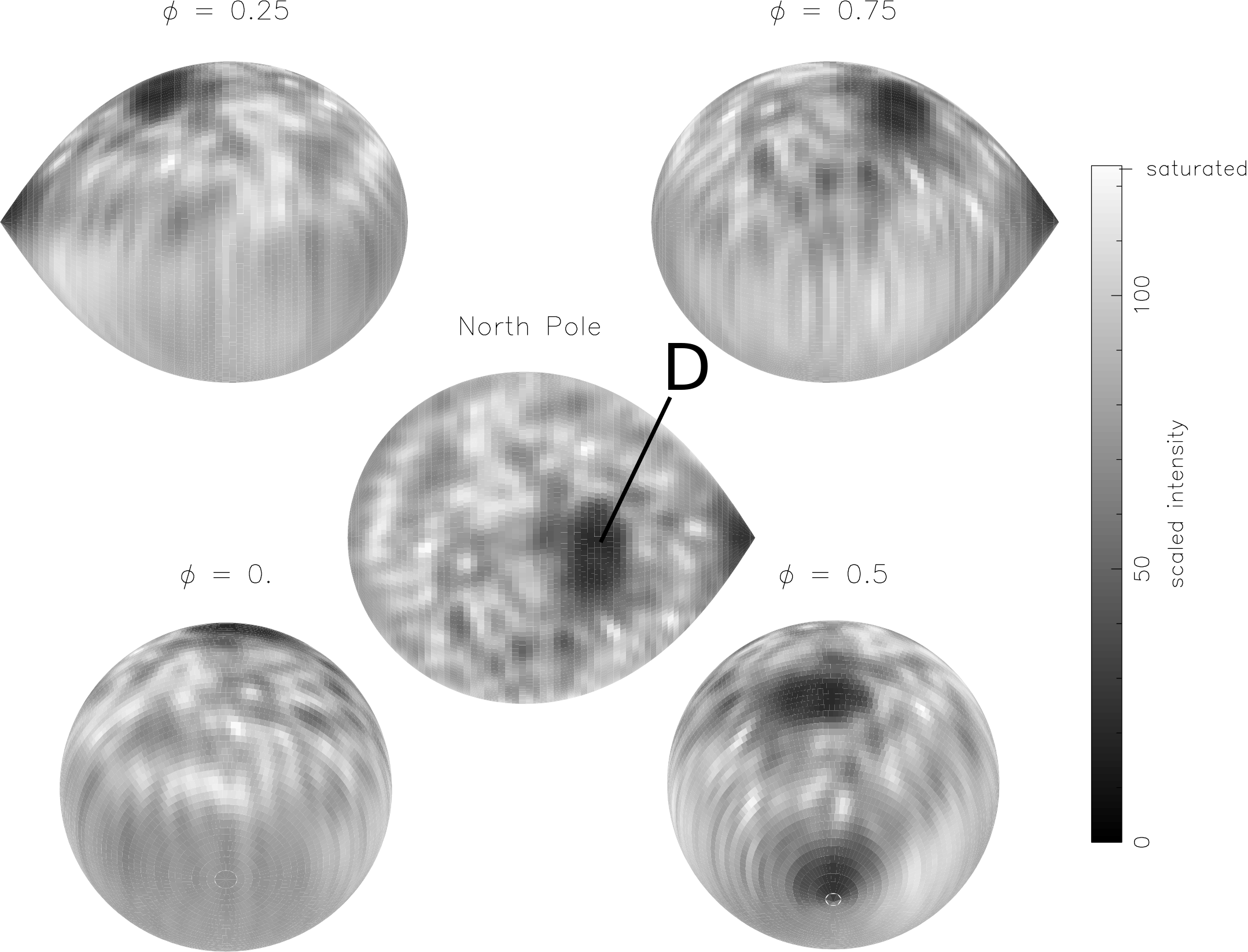}
\caption{The same as Figure~\ref{fig:map01} but for D05.}
\label{fig:map05}
\end{figure}

\begin{figure}
\centering
\includegraphics[width=0.5\textwidth]{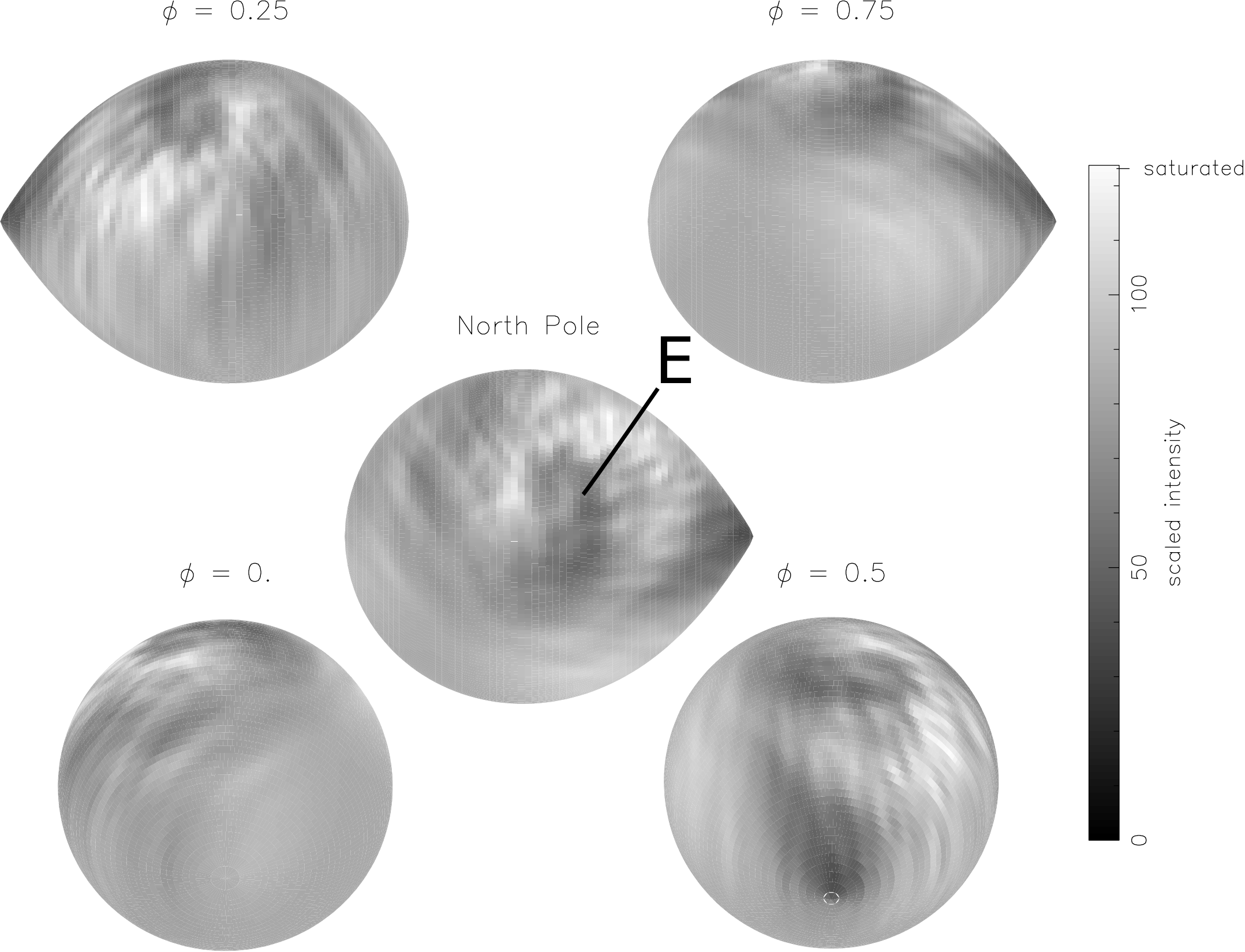}
\caption{The same as Figure~\ref{fig:map01} but for D06.}
\label{fig:map06}
\end{figure}

\begin{figure}
\centering
\includegraphics[width=0.5\textwidth]{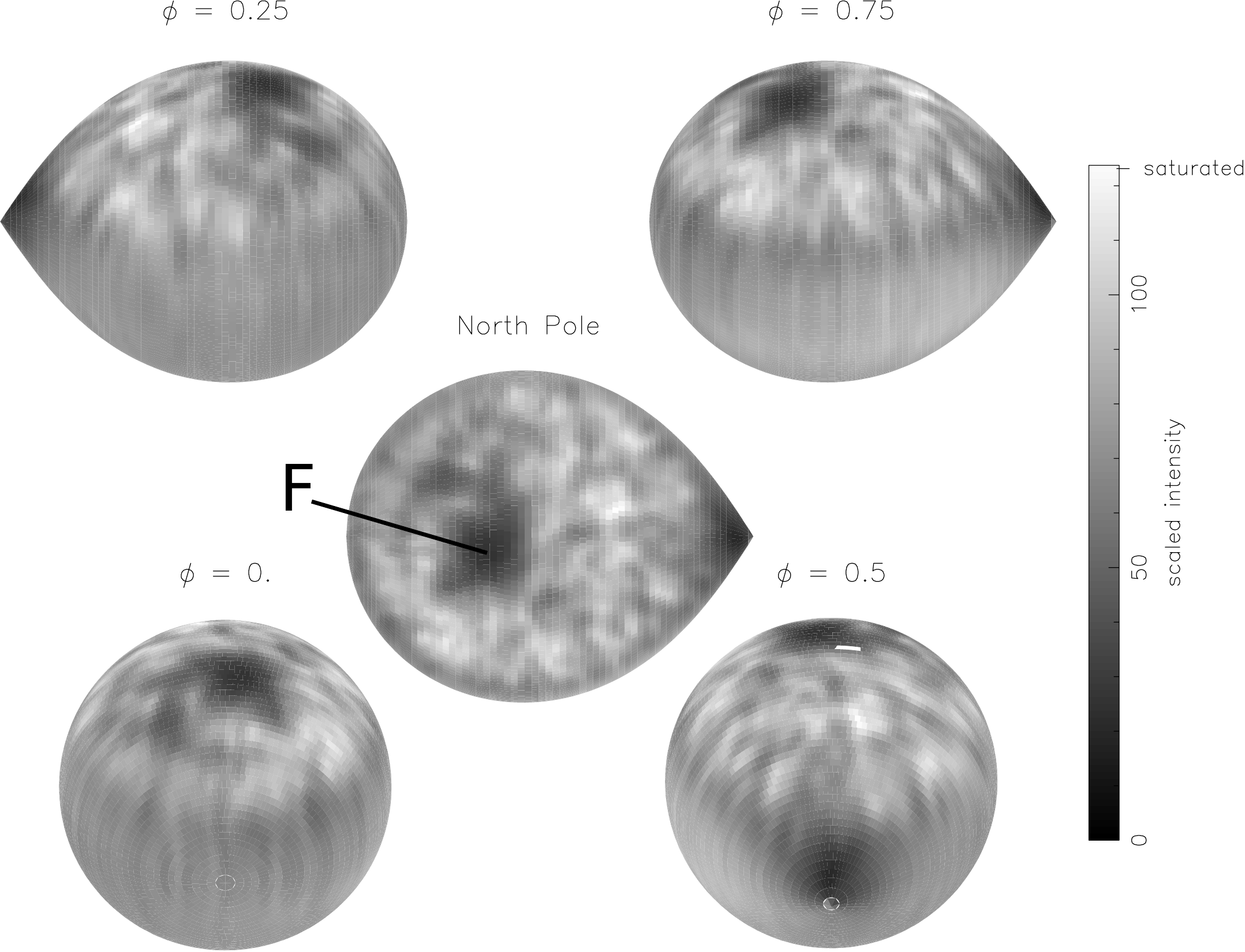}
\caption{The same as Figure~\ref{fig:map01} but for D08.}
\label{fig:map08}
\end{figure}

\begin{figure}
\centering
\includegraphics[width=0.5\textwidth]{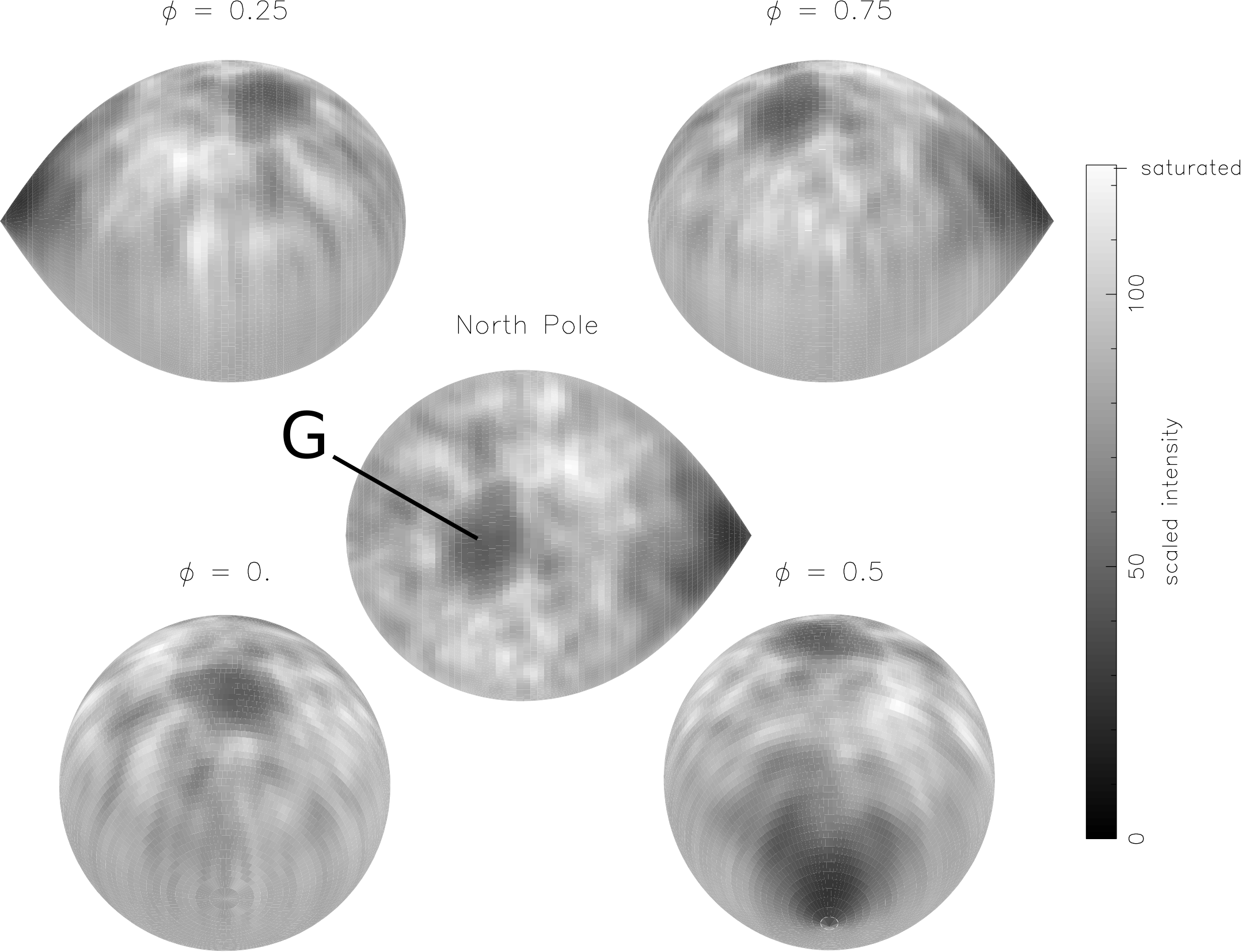}
\caption{The same as Figure~\ref{fig:map01} but for D09a.}
\label{fig:map09a}
\end{figure}

\begin{figure}
\centering
\includegraphics[width=0.5\textwidth]{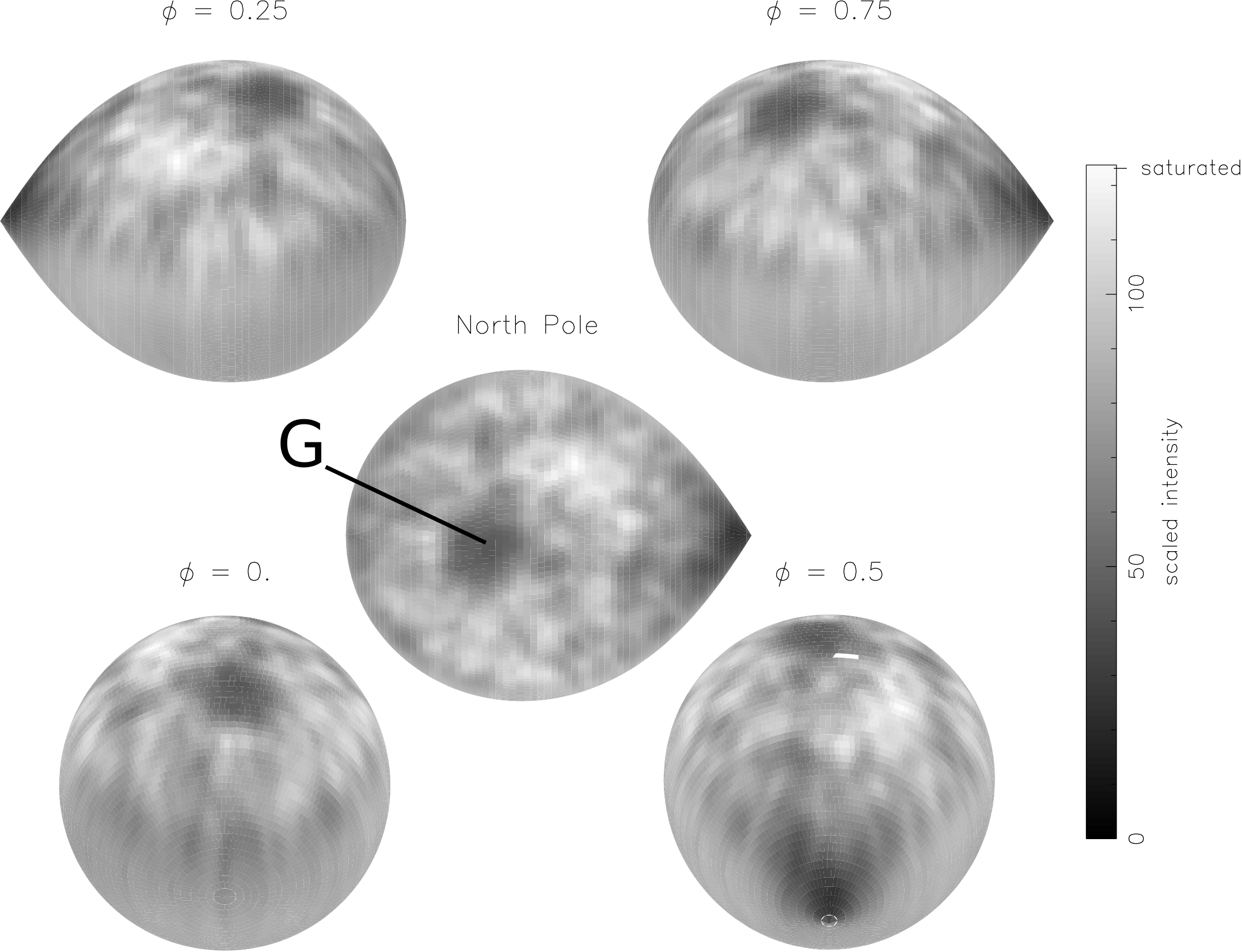}
\caption{The same as Figure~\ref{fig:map01} but for D09b.}
\label{fig:map09b}
\end{figure}

Roche tomograms of the secondary star in AE Aqr were constructed for each data set using the system parameters derived in Section~\ref{sec:incsystemicmasses}. The corresponding fits to the data are shown in Figures~\ref{fig:trails04}~to~\ref{fig:trails08}, and the Roche tomograms are shown in Figures~\ref{fig:map04}~to~\ref{fig:map08}. For ease of comparison, the previously published Roche tomograms of D01 and D09a~\&~D09b are shown in Figures~\ref{fig:map01},~\ref{fig:map09a}~\&~\ref{fig:map09b}, respectively. These were previously analysed by \cite{watson2006} and \cite{hill2014}, and we highlight the relevant features here. In the analysis which follows, the map coordinates are defined such that $0\degr$ longitude is the centre of the back of the star, with increasing longitude towards the leading hemisphere, and with the L$_{1}$ point at $180\degr$. We note that, due to the inclination of the system combined with limitations in the technique, we only consider features mapped in the Northern hemisphere to be reliable. Hence, the Southern hemisphere is excluded from our analysis.

We have adopted the optimal system parameters determined for each data set rather than the mean values across all data sets. This is due to the fact that systematic effects between data sets may result in different optimal system parameters being determined. Hence, adopting the mean values may lead to an increase in the number of artefacts reconstructed in the maps. Nevertheless, we have assessed the impact of adopting the mean system parameters by additionally carrying out the analysis in this section for Roche tomograms reconstructed using the mean values of $i = 57.4\degr$, $M_{1} = 0.92$~M$_{\sun}$ and $M_{2} = 0.61$~M$_{\sun}$ (where the component masses were calculated from the mean of the constant $M_{(1,2)}\sin^{3}{i}$). We find that, for each data set, the spot features reconstructed using the mean parameters are not significantly different to those reconstructed using the optimal parameters. Furthermore, there are no significant differences in the fractional spot coverage as a function of longitude and latitude (see Section~\ref{sec:spotcover}) on the maps reconstructed using the mean and the optimal parameters. This shows the robustness of the surface features reconstructed against incorrect inclination and component masses. However, to obtain the same map entropy using the mean parameters, as compared to the optimal parameters, we were required to fit the data to a higher aim $\chi^{2}$. Thus, we adopted the optimal parameters for the spot analysis as they provide a better fit to the data.

\subsection{Global features}
\label{sec:global}
Spot features, both large and small, are clearly prevalent in all tomograms. Common to all maps are the dark regions around the L$_{1}$ point, primarily due to the time-averaged effects of irradiation (ionising weak metal lines preventing photon absorption), as well as gravity darkening. As both of these effects appear dark in the maps, we are unable to clearly distinguish between the two. Likewise, we are unable to disentangle any starspots that may inhabit the affected region. Nevertheless, the impact of gravity darkening on the fractional spot coverage, \f, was assessed, and is discussed in Section~\ref{sec:spotcover}.

The map of D01 (see Figure~\ref{fig:map01}) clearly shows a single large spot (labelled `A') extending 60--80$\degr$~latitude and 260--320$\degr$~longitude, with a second prominent spot centred at 50$\degr$~latitude and $180\degr$~longitude. Also clearly apparent is a spot extending from $\sim40\degr$ latitude down to the L$_{1}$ point, becoming indistinguishable in the irradiated region. The map of D04 (see Figure~\ref{fig:map04}) shows two separate large spots, with the largest (labelled `B') extending 40--75$\degr$ in latitude and 160--220$\degr$ in longitude, and the second largest (labelled `C')  covering 60--70$\degr$~latitude and 310--340$\degr$~longitude. The map of D05 (see Figure~\ref{fig:map05})  has a single large spot (labelled `D'), extending 55--75$\degr$ in latitude and 140--210$\degr$ in longitude. The poor phase coverage of AE Aqr in D06 resulted in the reconstructed map having a relatively featureless leading hemisphere, with spots on the rest of the star becoming smeared out over the image, resulting in a lower contrast (see Figure~\ref{fig:map06}). Despite this, at least one, possibly two large spots (labelled `E') are evident above $\sim65\degr$ latitude, centred on $\sim180\degr$ longitude, with another prominent feature extending from $40\degr$ down to the L$_{1}$ point. The map of D08 (see Figure~\ref{fig:map08}) exhibits a single large spot (labelled `F') extending 65--85$\degr$ and 340--060$\degr$ in latitude and longitude, respectively. Finally, the maps of D09a and D09b (see Figures~\ref{fig:map09a}~and~\ref{fig:map09b}) show one large spot (labelled `G') spanning 65--90$\degr$~latitude and 340--050$\degr$~longitude.

The latitude of the largest spot in each map (labelled A, B, D, E, F, G) remains fairly constant over the 8~years between the first and last observations, ranging between 60--80$\degr$. However, the longitude of the dominant spot is not fixed. In D01, the dominant spot (A) lies $\sim290\degr$~longitude, whereas for D04, D05 and D06, the dominant spot (labelled B, D and E, respectively) lies $\sim180\degr$~longitude, although the position of spot E in D06 is less certain due to smeared features. In contrast, the largest spot in D08 (labelled F) and D09a~\&~D09b (labelled G) lies at $\sim15\degr$~longitude. Such monolithic spots are prevalent in many Doppler imaging studies of rapidly-rotating solar-type stars such as LQ~Hya \citep{donati1999}, and in CVs such as BV~Cen \citep{watson2007}. The high-latitude spots imaged here, and their possible evolution, are further discussed in Section~\ref{sec:discussion}.

Starspots are also prevalent at low to mid latitudes in all maps, and in order to make a more quantitive assessment of their properties and the underlying dynamo mechanism, we must consider their size and distribution across the stellar surface.

\subsection{Pixel intensity and spot filling factor} \label{sec:intensity}
\begin{figure}
\centering
\includegraphics[width=0.5\textwidth]{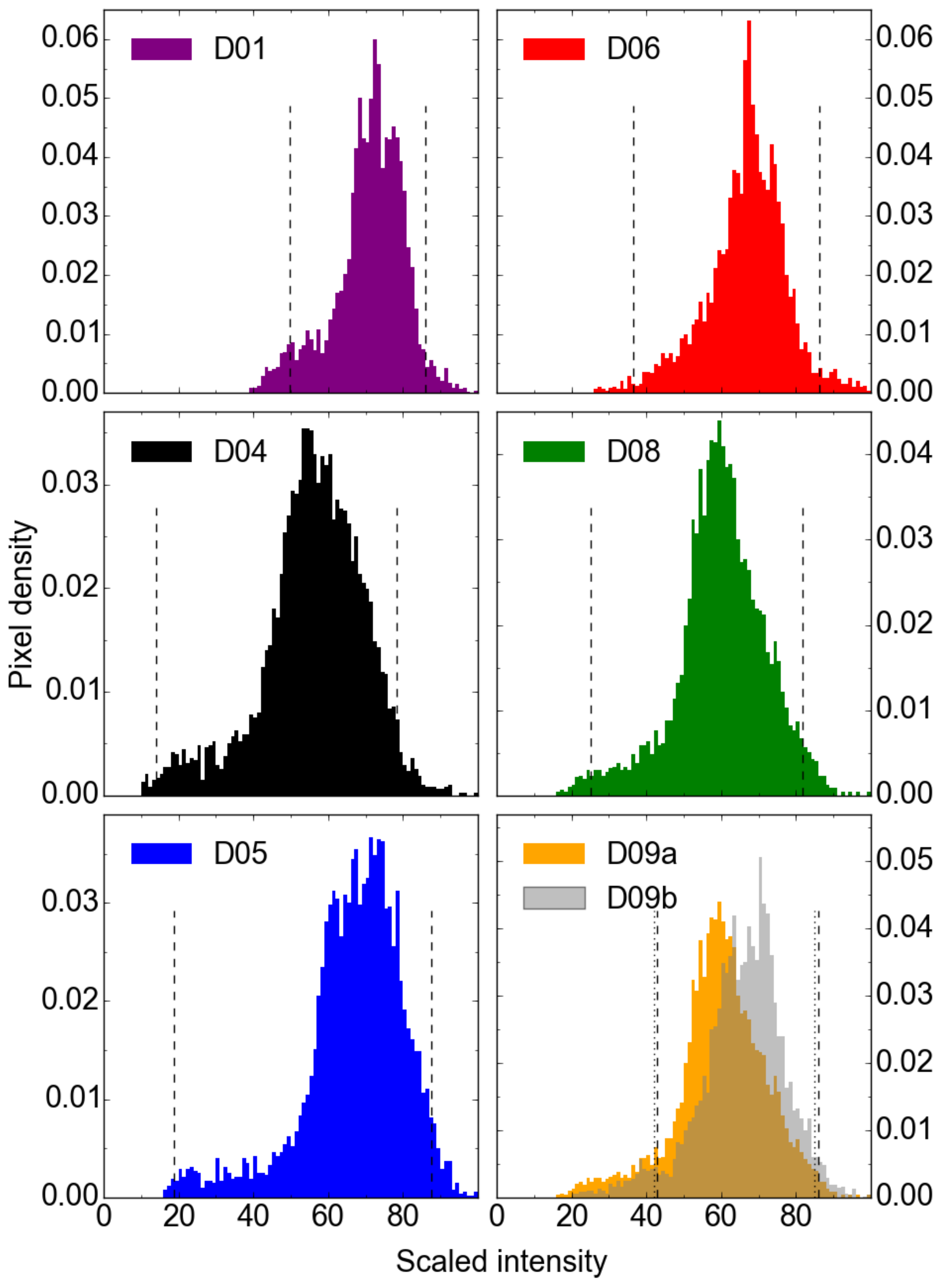}
\caption{Histograms of the pixel intensities in the Roche tomograms of AE~Aqr for each data set, where the pixel density is the fraction of the total number of pixels. Pixels on the Southern hemisphere (latitude $<0\degr$) were not included (see Section~\ref{sec:intensity} for details). The brightest pixel in each map was assigned an intensity of 100 and all other pixel intensities were scaled linearly against this. The definition of the pixel intensity representing the immaculate photosphere is shown as a dashed line on the right side of the histogram (a dotted line for D09b), and that representing a totally spotted pixel is shown as a dashed line on the left side (see text for details).}
\label{fig:histogram}
\end{figure}

To determine the spot coverage in the Roche tomograms, it was first necessary to define the pixel intensity of the immaculate photosphere as well as that of a totally spotted pixel. We do not adopt a two-temperature model when fitting with Roche tomography, where a spot filling factor is predetermined (e.g. \citealt{cameron1994}), as secondary stars in CVs are expected to exhibit large temperature differences due to irradiation by the primary. Our method of determining a totally spotted pixel was to simply select the lowest pixel intensity at the centre of the largest spot feature. The adopted value of a totally spotted pixel for each map is shown as a dotted line on the left side of the histograms of pixel intensities in Figure~\ref{fig:histogram}, where the brightest pixel is assigned an intensity of 100 and other pixels scaled linearly relative to this. Pixels with a lower intensity are present in the Roche tomograms, but these are confined to the region around the L$_{1}$ point, and as discussed above, are not likely to be due to a spot feature. Pixels on the Southern hemisphere are not included in the histograms or any of the analysis below for two reasons. Due to the inclination of the binary, a large portion of the surface is not visible, and so a substantial number of pixels on the Southern hemisphere are assigned the default map value. Furthermore, as RVs cannot constrain whether a feature is located in the Northern or Southern hemisphere, features may be mirrored about the equator, reducing their contrast as they are smeared over a larger area.

The intensity of the immaculate photosphere was more difficult to define due to the growth of bright pixels -- an artefact known to affect maps that are not thresholded (e.g. \citealt{hatzes1992}). To assess the extent of bright pixel growth in our Roche tomograms, we carried out reconstructions of simulated maps with a random spot distribution, adding in varying levels of noise. We found that the brightest reconstructed pixel was between 4--15~per~cent higher in intensity than that of the original map, and that up to 12\% of pixels were classed as `bright' (typically $\sim1$~per~cent). Data with a higher SNR increased the number of bright pixels, however, the most dramatic increase was found when the data were fit to progressively lower reduced~$\chi^{2}$. Indeed, the maps with a large number of bright pixels were clearly overfit, exhibiting substantial reconstructed noise. Given that our Roche tomograms of AE~Aqr were fit to an aim reduced~$\chi^{2}$ that limited the amount of reconstructed noise (see Figure~\ref{fig:chivsentropy}), we assumed a percentage growth of bright pixels of 3\% of the total number of pixels in the map. This somewhat conservative estimate, given the results of our simulations, means we likely underestimate \f. Hence, we defined the immaculate photosphere as the lowest pixel intensity that includes 97\% of all pixels, and is shown in Figure~\ref{fig:histogram} as a dotted line on the right side of the histograms. 

The histograms of pixel intensities in Figure~\ref{fig:histogram} show broad peaks, with long tails towards lower intensities. An idealised histogram would have a significantly bimodal distribution of pixel intensities, where spotted pixels and those representing the immaculate photosphere would be more clearly separated. The large number of intermediate pixel intensities found here may be explained by a lack of contrast in the maps, stemming from both a population of unresolved spots, as well as spots that have been smeared in latitude, increasing their areal coverage.

\subsection{Spot coverage as a function of longitude and latitude}
\label{sec:spotcover}
Figures~\ref{fig:spotcoverlong}~and~\ref{fig:spotcoverlat} show the fractional spot coverage \f as a function of longitude and latitude for each map, respectively. We calculate \f using Equation~\ref{eqn:fsc}, where $I$ is the pixel intensity, $I_{\mathrm{p}}$ is the intensity of the immaculate photosphere, and $I_{\mathrm{s}}$ is the intensity of a totally spotted pixel. 

\begin{equation}
\f = \text{max}\left[0,\text{min}\left[1, \frac{I_{\mathrm{p}} - I}{I_{\mathrm{p}} - I_{\mathrm{s}}}\right]\right]
\label{eqn:fsc}
\end{equation}

The most prominent feature common to all plots is the large value of \f around $180\degr$ longitude in Figure~\ref{fig:spotcoverlong}, as well the increased coverage below $20\degr$ latitude in Figure~\ref{fig:spotcoverlat}. These regions include the features around the L$_{1}$ point which are dominated by the effects of irradiation and gravity darkening (as discussed above). The impact of these phenomena on the maps was assessed by simulating blank maps using the corresponding system parameters and limb darkening coefficients specific to each data set. Additionally, a gravity darkening coefficient of $\beta = 0.1$ was adopted, as this is representative of that measured for late-type secondary stars in close binaries \citep{djurasevic2003,djurasevic2006}. The simulated maps were used to create synthetic line profiles with the same orbital phases, exposure times and instrumental resolutions of the original data. These synthetic line profiles were then reconstructed in the same manner as the original data, and \f was calculated using the same definitions of a totally-spotted pixel and that representing the immaculate photosphere, as determined above. The value of \f of these synthetic maps was then subtracted from that of the original maps, effectively removing the systematic effects of inclination, phase sampling, and incorrect limb and gravity darkening in the spot coverage of the original maps. However, we cannot distinguish between a spot and the effects of gravity darkening or irradiation, as all three appear as dark regions in the maps. Hence, when we apply the correction described above, spots that are located in the regions most affected by irradiation and gravity darkening are not preserved (as the regions are made brighter, regardless of spots being present; see discussion in Section~\ref{sec:global}), and thus we may underestimate \f in those regions (see Figures~\ref{fig:spotcoverlong}~and~\ref{fig:spotcoverlat}).

\subsubsection{Longitude cover}
\label{sec:longcover}
\begin{figure}
\centering
\includegraphics[width=0.5\textwidth]{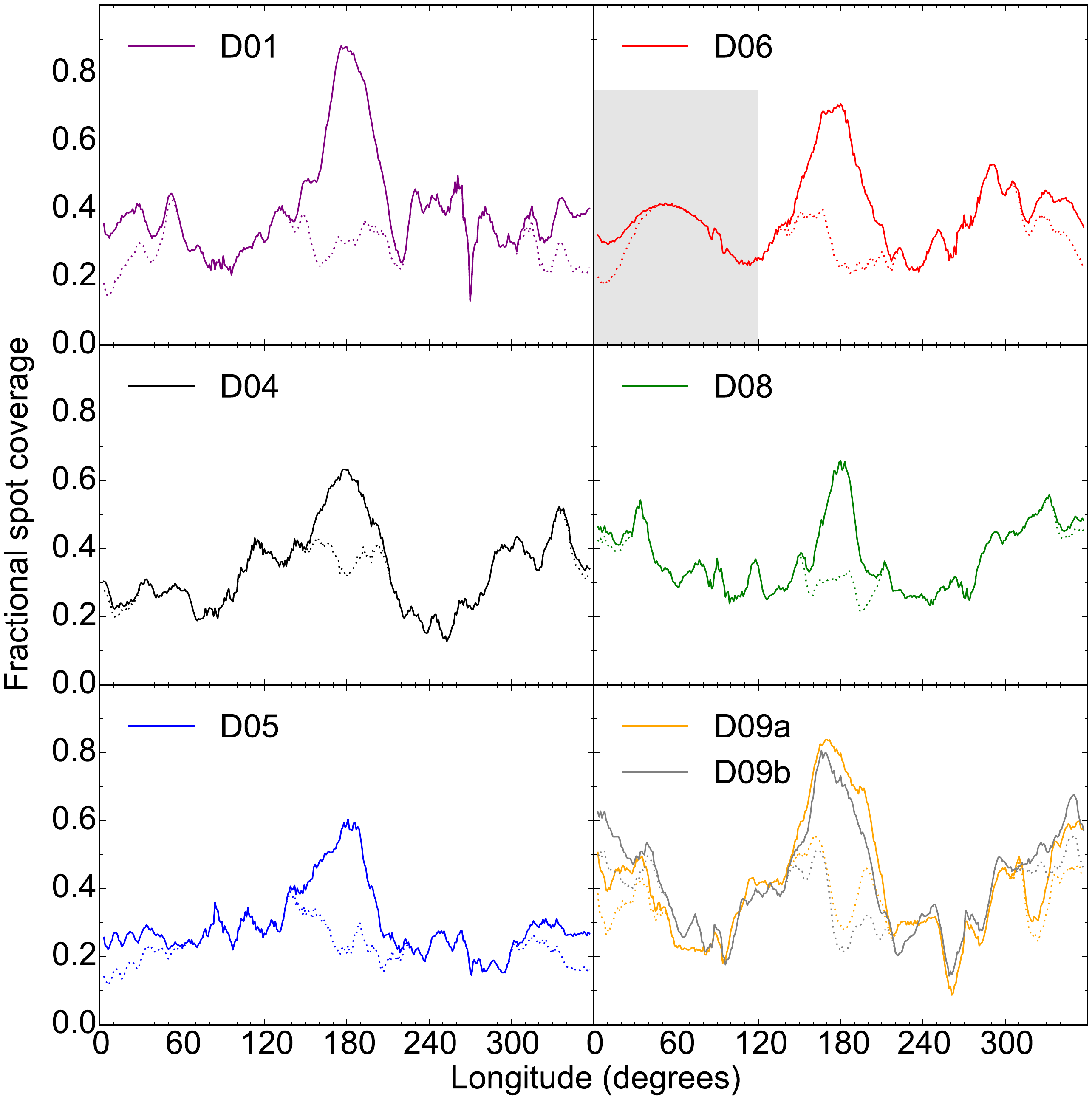}
\caption{The fractional spot coverage \f as a function of longitude for the Northern hemisphere of AE~Aqr for each data set, where \f is normalized by the number of pixels within a $6\degr$~longitude bin. The solid line shows \f for the original map, and the dotted line shows \f after subtracting \f for the simulated map (removing the effects of irradiation and gravity darkening, see Section~\ref{sec:spotcover} for details). The shaded region for D06 indicates that features in this region are not reliable due to a lack of phase coverage.}
\label{fig:spotcoverlong}
\end{figure}

The fractional spot coverage \f as a function of longitude varies significantly for each map (see Figure~\ref{fig:spotcoverlong}). Even after subtraction of the simulated maps there still remains a significant spot coverage around $180\degr$~longitude for D04, D05, D06 and D09a~\&~D09b. The map of D01 shows an increase in \f around $250\degr$~longitude, with the maps of D04 and D06 showing a significant increase in \f between 280--360$\degr$~longitude, and the maps of D08 and D09a~\&~D09b showing a larger \f over a broader range of 280--060$\degr$~longitude. The increase in \f between 0--120$\degr$~longitude for D06 is not real and results from poor phase coverage, leading to the pixels in this region being assigned a value similar to that of the default map (the mean of all map pixels). 

The distributions of \f in longitude suggests the existence of two longitude regions, separated by $\sim180\degr$, with significantly higher \f in the maps of D04, D06 and D09a~\&~D09b. However, no such distributions are present in D01 and D05, with the high \f around $180\degr$ in D08 becoming much lower once the effects of irradiation and gravity darkening have been removed. Such `active longitudes' are observed in single stars such as LQ~Hya, AB~Dor and EK~Dra \citep{berdyugina2005} as well as in RS CVn binaries (e.g. \citealt{rodono2000}). In shorter period systems ($\prot<1$~d), active regions are preferentially located at quadrature longitudes (e.g. \citealt{olah1994,heckert1998}), whereas longer period systems show no such preference. Indeed the RS CVn binaries HD~106225 \citep{strassmeier1994} and II~Peg \citep{rodono2000} show a migration of active longitudes with respect to the companion star, ascribed to differential rotation. However, the lack of clearly defined active longitudes in the maps of D01, D05 and D08 suggest that such active regions are not permanent features in AE~Aqr, or are at least not fixed with respect to the companion. We discuss possible explanations of the observed stellar activity in Section~\ref{sec:discussion}.

\subsubsection{Latitude cover}
\label{sec:latcover}
\begin{figure*}
\centering
\includegraphics[width=\textwidth]{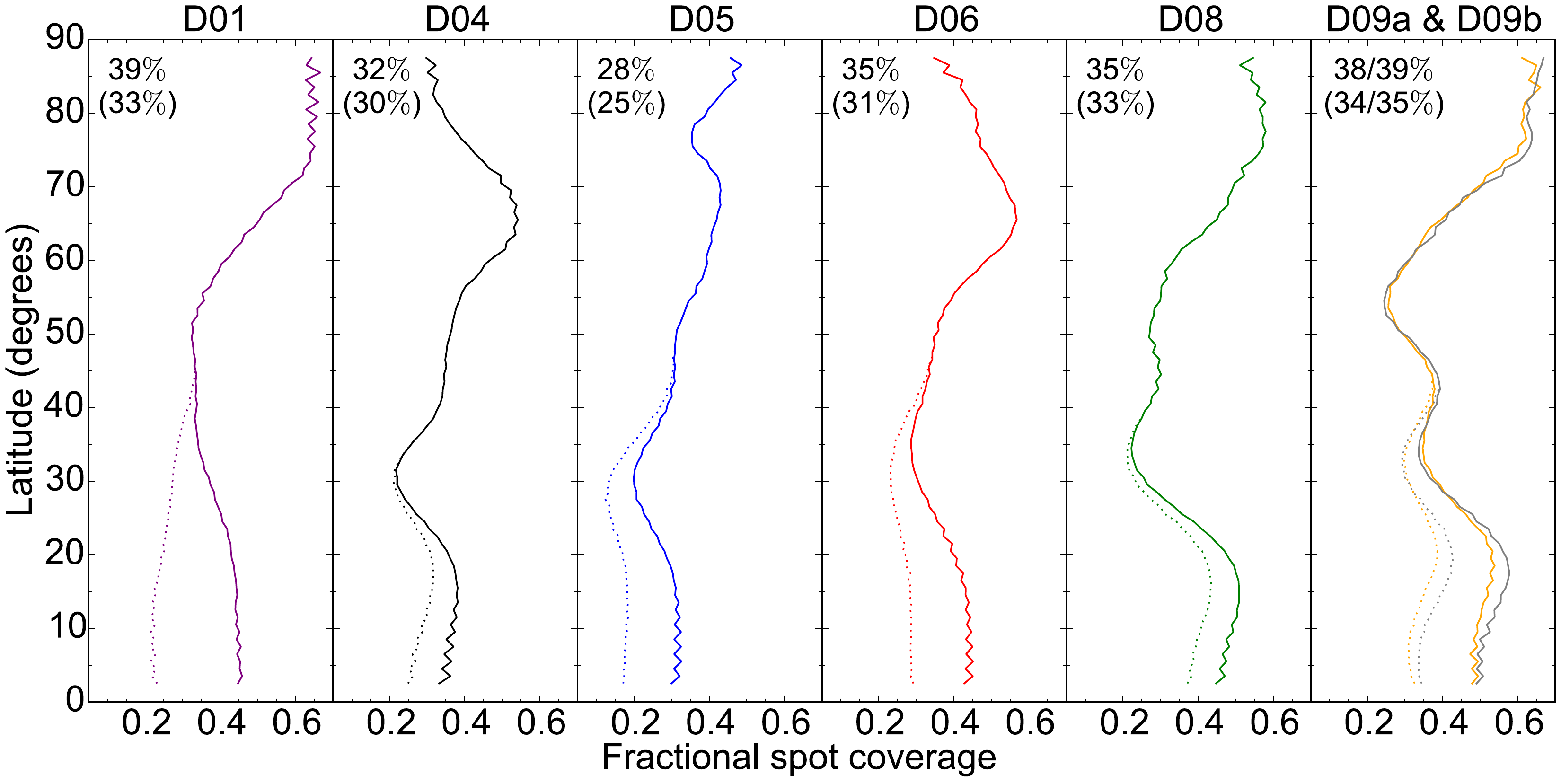}
\caption{The fractional spot coverage \f as a function of latitude for the Northern hemisphere of AE~Aqr for each data set, normalized by the surface area at that latitude. The solid line shows \f for the original map, and the dotted line shows \f after subtracting \f for the simulated map (removing the effects of irradiation and gravity darkening, see Section~\ref{sec:spotcover} for details). The total spot coverage for each map is given as a percentage in the top left of each panel, where the value in parenthesis is the total spot coverage after subtracting the simulated map.}
\label{fig:spotcoverlat}
\end{figure*}

The fractional spot coverage \f as a function of latitude is shown in Figure~\ref{fig:spotcoverlat} for all maps. Latitudes above $60\degr$ show high \f due to the large high-latitude spots. The distribution of \f above $60\degr$~latitude is similar in D01, D08, and D09a~\&~D09b, with maximum \f at latitudes $>70\degr$. In comparison, \f peaks around $65\degr$~latitude for D04 and D06, with the flatter distribution of D05 exhibiting a peak that is consistent with D04 and D06.

At mid-latitudes of $\sim45\degr$ we see the apparent growth of \f over the 8~years of observations, with a peak becoming more pronounced in the plots of D08 and D09a~\&~D09b. This may be indicative of increasing spot coverage at this latitude, however, it could also be explained by a relative \emph{decrease} in the spot coverage at surrounding latitudes. Given our lack of an absolute spot filling factor, the total spot coverages between maps (given as a percentage in the top-left of each panel in Figure~\ref{fig:spotcoverlat}) are not necessarily comparable on an absolute scale, rather, they indicate the relative spot coverage for that particular map. However, given the measured spot coverage increases between D05 and D09b, we can be confident the increase of \f at $45\degr$~latitude is indeed due to an increase in spot coverage localised to this latitude. 

A common feature to all maps, excluding that of D01, is the apparent lack of spots around $30\degr$~latitude. This is most obvious for D04, D05 and D08, as \f is lowest at this latitude, and becomes more pronounced for D06 and D09a~\&~D09b when the systematic effects of irradiation and gravity darkening are subtracted. The lower quality of data in D01 means features at lower latitudes are more smeared out compared with other maps in our sample, and so while this feature may exist, we cannot resolve it to the same degree as in the other maps.

At lower latitudes, we see a clear increase in \f around $20\degr$ for all maps. As previously discussed, the effects of irradiation and gravity darkening are significant at these latitudes, which makes it difficult to determine how \f varies between maps. However, even after these effects have been subtracted, we still see a persistently high-level of \f around $20\degr$~latitude, suggesting spots at this latitude are common.

The reliability of the reconstruction of small scale features in the maps was tested with reconstructions of simulated data sets. Test maps were created using the same parameters as those for D04, D05 and D08, with a large polar spot, bands of spots at $45\degr$ and $20\degr$ latitude, and gravity and limb darkening. The bands of spots each contained at least 17 individual spots, with sizes ranging between 5--10$\degr$ in latitude and longitude, separated by 10--20$\degr$ longitude. Trailed spectra were created using the same phases as the original data, and representative noise was added. Maps were reconstructed from the synthetic data in the same manner as that carried out for the original data. The resulting maps of all three simulated data sets show that the polar spot and the spot band at $45\degr$ are clearly recovered, with the spot band at $20\degr$ becoming moderately smeared in latitude, reducing the contrast. However, the latter spots are still clearly distinct from the reconstructed noise, and so we are confident that we can reliably reconstruct features of this size at these latitudes. Furthermore, the latitudinal spread in \f due to these smeared spots has been taken into account in the analysis presented here.

\section{Discussion}
\label{sec:discussion}

\subsection{Evolution of high-latitude spots}
\label{sec:highlatspots}
Several high-latitude spots are seen in the maps of AE~Aqr, and are labelled A-G in Figures~\ref{fig:map01}-\ref{fig:map09b}. There are several possible scenarios that may explain the behaviour of the dominant high-latitude spot; The largest spot in each map may be the same feature, with only its position changing between observations.
Alternatively, the largest spot may not be the same feature, but instead is evolving significantly, disappearing and appearing elsewhere between maps.

In the first scenario, if the largest spot in each map (labelled A, B, D, E, F, G) is the same feature, and is able to move position, then stable spots may have a lifetime of $\sim8$~yr. However, if the largest spot must remain in a fixed position, then spots B, D and E in Figures~\ref{fig:map04}--\ref{fig:map06} imply that a large, stable spot can live for $\sim4$~yr. The changing position of the dominant spot may be explained by the presence of differential rotation (DR) on the surface of the star, as measured by \cite{hill2014} using the maps of D09a~\&~D09b. By using the measured shear-rate, we have determined the longitude-shift for latitudes between 60--80$\degr$ over the intervals between observations. By comparing the shift in longitude due to DR with the observed change in longitude of the spot in question, we find that spot~A in D01 may have shifted to the position of spot~B in D04, and spot~E in D06 may have shifted to the position of spot~F in D08. Only the movement of the spots in these two cases may be explained by DR, with all other observed shifts being incompatible with this mechanism. It is possible that the spot~C in D04 grew in area \emph{and} rotated due to DR to the same position as spot~D in D05, however, the spot~B in D04 would then have to disappear completely in D05, which seems unlikely. Furthermore, the fact that spot~F in D08 is at the same location as spot~G in D09a~\&~D09b suggests DR is unlikely to be the cause of the observed shifts. Indeed DR may primarily affect smaller spots that have magnetic flux tubes anchored closer to the surface, whereas larger spots are less affected as their flux tubes may be anchored deeper in the stellar interior. 

The second scenario requires the evolution of these dominant spots over a relatively short time period. Similar behaviour, observed in other systems, has been explained by a phenomenon known as the `flip-flop' activity cycle. In this scenario, active longitudes are present on the star. The `flip-flop' cycle occurs when the active longitude with the highest level of activity (i.e. the most spotted) switches to the opposite longitude, with cycles taking years to decades to occur (see \citealt{berdyugina2005} for a summary). The disappearance of spot~C in D04, as compared to D05, may imply the switching of dominant longitudes between observations of AE~Aqr. However, the lack of at least two clearly defined active longitudes in D01, D05 and D08 suggests that this type of cycle is not present in AE~Aqr. In any case, a robust detection of a `flip-flop' cycle in AE~Aqr would require a much shorter gap between observations than obtained for our sample in order to clearly track the emergence of spots at preferred longitudes. 

\subsection{Magnetic flux tube dynamics in close binaries}
\label{sec:magfluxtube}
To understand why surface features are distributed as they are, and why they evolve in the observed manner, we can compare the Roche tomograms to numerical simulations of emerging magnetic flux tubes in close binary systems. \cite{holzwarth2003a} carried out such simulations, assuming that starspots are formed by erupting flux tubes that originate from the bottom of the stellar convective zone (by analogy with the Sun). Magnetic fields, believed to be amplified in the rotational shear layer (tacholine) near the base of the convection zone, are stored in the form of toroidal flux tubes in the convective overshoot layer \citep{schussler1994}. By studying the equilibrium and linear stability properties of these flux tubes, \cite{holzwarth2003a,holzwarth2003b} examined whether the influence of the companion star was able to trigger rising flux loops at preferred longitudes, since the presence of the companion breaks the rotational symmetry of the star. 

The authors find that while the magnitudes of tidal effects are rather small, they nevertheless lead to the formation of clusters of flux tube eruptions at preferred longitudes on opposite sides of the star, a phenomenon resulting from the resonating action of tidal effects on rising flux tubes. Pertinently, the authors establish that the longitude distribution of spot clusters on the surface depends on the initial magnetic field strength and latitude of the flux tubes in the overshoot region, implying there is no preferred longitude in a globally fixed direction. Moreover, flux tubes that are perturbed at different latitudes in the convective overshot region, show a wide latitudinal range of emergence on the stellar surface, with considerable asymmetries appearing as highly peaked \f distributions or broad preferred longitudes. In a binary with $P_{\text{orb}} = 2$~d, the authors find it takes several months to years for a flux tube, perturbed from the convective overshoot region, to emerge on the stellar surface. Over this time, the Coriolis force acts on the internal gas flow and causes the poleward deflection of the tube \citep{schussler1992}, with the largest deflection for flux tubes starting at lower latitudes, and those starting $>60\degr$~latitude showing essentially no deflection. Such simulations are consistent with the large high-latitude spots found on AE~Aqr.

Clearly the behaviour of flux tubes in a binary system is complex. However, the results of the models by \cite{holzwarth2003b} may explain the varying size, distribution and evolution of the starspots imaged in AE~Aqr; Namely, the low value of \f around $30\degr$~latitude in the maps of AE~Aqr is consistent with the fact that some latitudes are avoided by the simulated erupting flux tubes. Furthermore, the variable peak of \f at high latitudes in AE~Aqr may be a redistribution of magnetic energy, changing the field strength of perturbed flux tubes and causing them to emerge at different latitudes, as well as shifting in longitude. In addition, flux tube eruption at high latitudes, due to flow instabilities, leads to spots emerging over a broad longitude region -- similar to the large high-latitude spots observed in AE~Aqr.

\section{A magnetic activity cycle?}
It is unclear what the dominant dynamo mechanism is in AE~Aqr. There is no clear evidence that we see a `flip-flop' cycle, especially given the lack of active longitudes in three maps of our sample. Indeed the most prominent evidence of an activity cycle is the increase in \f at $45\degr$~latitude over the course of the 8~years of observations, combined with the persistently high \f around $20\degr$~latitude. The growth in spot coverage around $45\degr$~latitude may be indicative of an emerging band of spots, forming part of an activity cycle similar to that seen on the Sun. Furthermore, the increase in \f around $20\degr$~latitude may be a second band of spots that form part of a previous cycle. In the case of the Sun, the latitude of emergence of flux tubes gradually moves towards the equator over the course of an activity cycle, taking $\sim11$~years, with little overlap between consecutive cycles of flux tube emergence. However, simulations by \cite{isik2011} show that stronger dynamo excitation may cause a larger overlap between consecutive cycles. Given that the high spot coverage in AE~Aqr suggests a strong dynamo excitation, the presence of two prominent bands of spots may be indicative of such an overlap between cycles. However, if such a cycle were to exist, we would expect to see the higher-latitude peak move towards lower latitudes over the course of our observations. Given that we do not clearly see this, any solar-like activity cycle must take place over a timescale longer than 8~years. \cite{saar1999}, in their study of a large sample of stars in a range of systems (including single stars and binaries), found several correlations between the duration of the magnetic activity cycle and the rotation period. Pertinently, using the fit to all stars in their sample, we estimate that AE~Aqr would have a magnetic activity cycle lasting $\sim16$~years. Furthermore, we estimate a longer cycle period of $\sim22$~years by using the fit to stars defined as `superactive' by \cite{saar2001}. If the correlations found for other systems are also true for AE~Aqr, then we may have observed less than half of an activity cycle. 

This is the first time the number, size and distribution of starspots has been tracked in a CV secondary. While any specific interpretation of the long term behaviour of the imaged starspots is somewhat challenging, the presence and evolution of two distinct bands of spots may indicate an ongoing magnetic activity cycle in the secondary star in AE~Aqr. Hence, it is crucial we continue our study of its magnetic activity. Future maps would allow us to track the evolution of the large high-latitude spots to determine if their long term behaviour is periodic. Moreover, by tracking the evolution of the spot bands we may determine if they form part of a periodic activity cycle, and if so, the length of such a cycle could be measured, providing a unique insight into the behaviour of the stellar dynamo in an interacting binary. In addition, shorter, more intensive campaigns would allow the position of specific spot features to be tracked. This would allow further measurements of differential rotation as well as determining if meridional flows are present. The impact of tidal forces on magnetic flux tube emergence and possible quenching of mass transfer could then be assessed.

\section{Conclusions}
\label{sec:conclusions}
We have imaged starspots on the secondary star in AE~Aqr for 7~epochs, spread over 8~years. This is the first time such as study has been carried out for a secondary star in a CV and, in some cases, the number, size and distribution of spots varies significantly between maps. In particular, the changing positions of the large high-latitude spots cannot be explained by differential rotation, nor by the `flip-flop' activity cycle. At lower latitudes, we see the emergence of a band of spots around $45\degr$~latitude, as well as a persistently high spot coverage around $20\degr$~latitude. These bands may form part of an activity cycle similar to that seen in the Sun, where magnetic flux tubes emerge at progressively lower latitudes throughout a cycle. Furthermore, the complex distribution and behaviour of spots may be attributed to the impact of the companion WD on flux tube dynamics.

\section*{Acknowledgments}
We thank Tom Marsh for the use of his \textsc{molly} software package in this work, and VALD for the stellar line-lists used. We thank the staff at Carnegie Observatories for their assistance with the MIKE pipeline and for access to the Henrietta Swope Telescope. C.A.H. acknowledges the Queen's University Belfast Department of Education and Learning PhD scholarship, C.A.W. acknowledges support by STFC grant ST/L000709/1, and D.S. acknowledges support by STFC grant ST/L000733/1. This research has made use of NASA's Astrophysics Data System and the Ureka software package provided by Space Telescope Science Institute and Gemini Observatory. 

\bibliographystyle{mn_new}
\bibliography{references.bbl}
\label{lastpage}

\end{document}